\RequirePackage{ifpdf}
\documentclass[12pt]{article}
\usepackage{xcolor}
\usepackage{amsmath, amsfonts, amsthm, amssymb, graphicx}
\usepackage{xspace}
\usepackage{epsfig}
\usepackage{psfrag}
\usepackage{geometry}
\usepackage{colonequals}
\usepackage{hyperref}
\usepackage{tikz}
\usepackage[capitalise]{cleveref}
\hypersetup{colorlinks=true,citecolor=black,linkcolor=black,urlcolor=black}

\pdfoutput=1

\newcommand{\op}{\ensuremath{\mathcal{O}}\xspace}
\newcommand{\del}{\partial} 

\newcommand{\bra}[1]{\ensuremath{\langle #1 |}\xspace}
\newcommand{\ket}[1]{\ensuremath{| #1 \rangle}\xspace}

\newcommand{\vev}[1]{\ensuremath{\langle #1 \rangle}\xspace}

\newcommand{\du}[2]{_{ #1 }^{\phantom{ #1 } #2 }}
\newcommand{\ud}[2]{^{ #1 }_{\phantom{ #1 } #2 }}

\newcommand{\hf}{\frac{1}{2}}

\newcommand{\FF}{\mathcal F}
\newcommand{\GG}{\mathcal G}

\let\a=\alpha \let\b=\beta \let\g=\gamma \let\d=\delta \let\e=\epsilon
  \let\q=\theta  \let\k=\kappa
 \let\m=\mu \let\n=\nu  
\let\s=\sigma \let\t=\tau    
  \let\D=\Delta  
    
    \let\G=\Gamma

\newcommand{\bxx}[1]{\begin{#1}}
\newcommand{\be}{\bxx{equation}}
\newcommand{\ee}{\end{equation}}

\def\ea{\end{array}}

\numberwithin{equation}{section}

\begin{document}
\title{\textbf{Counterterms in Truncated Conformal Perturbation Theory}}
\author{Daniel Rutter, Balt C. van Rees \\\\ \textsl{\small Centre for Particle Theory, Department of Mathematical Sciences}\\
\textsl{\small Durham University, DH1 3LE, UK}
}
\maketitle
\vskip 2cm
\begin{abstract}
We investigate the perturbative renormalisation of deformed conformal field theories from the Hamiltonian perspective. We discuss the relation with conformal perturbation theory, to which we provide an explicit match up to third order in the coupling, and show how second-order anomalous dimensions in the Wilson-Fisher fixed points are straightforwardly computed in the Hamiltonian framework. The second part of the paper focuses on the cutoff employed in the truncated conformal space approach of Yurov and Zamolodchikov \cite{Yurov:1989yu}. We discuss the appearance of non-covariant and non-local counterterms to second order in the cutoff, which we concretise in the $\phi^4$ theories, and find a smooth cutoff to deal with subleading oscillations.
\end{abstract}
\newpage

\tableofcontents
\newpage


\section{Introduction}
In this work, we address several questions about perturbative renormalisation from the Hamiltonian perspective. Our main interest in this method is its importance for the truncated conformal space approach (TCSA) of Yurov and Zamolodchikov \cite{Yurov:1989yu}. We will illustrate our general results with the $\phi^4$ theory in $d$ dimensions. 

The TCSA is the Rayleigh-Ritz method adapted to quantum field theory. The main idea is to truncate the discrete Hilbert space of a field theory on a compact spatial manifold to a certain finite-dimensional vector space. This in particular truncates the Hamiltonian to a finite-dimensional matrix, which one then proceeds to diagonalize numerically to obtain an estimate of the field theory spectrum. In practical computations one is confronted with an exponential growth in the number of states which necessitates improvements to this `bare' procedure in order to obtain meaningful results. One such an improvement is to add counterterms to the Hamiltonian in order to approximately take into account the effect of states above the cutoff. This idea was introduced first in \cite{Feverati:2006ni} and implemented and refined in several other works: see \cite{Watts:2011cr,Lencses:2014tba,Giokas:2011ix,BERIA2013457,Toth:2006tj} and more recently \cite{PhysRevD.91.025005} (which includes a review of earlier works) and \cite{Rychkov:2014eea,Rychkov:2015vap,Elias-Miro:2015bqk,Elias-Miro:2017tup,Elias-Miro:2017xxf}; a recent review is \cite{James:2017cpc}.

Our first question concerns the connection between the anomalous dimensions of composite operators in the plane, and the eigenvalues of the Hamiltonian on the cylinder. For conformal field theories (CFTs), the state-operator correspondence dictates that these ought to be completely equivalent, and whilst this is easily verified at first order in perturbation theory (see e.g. \cite[Chapter 5]{Cardy.ScalingAndRenormalization}), it becomes less straightforward at the next order. Here, we find the explicit relation up to third order using an argument which is easily extendable to higher orders. In \cref{sec:hamvslag}, we explain that the precise connection is provided by using conformal perturbation theory on the plane, rather than the usual Feynman diagrams. In \cref{sec:scalarthy}, we will use these newfound equations to compute the anomalous dimensions at the Wilson-Fisher fixed points to second order in the epsilon expansion. This computation is remarkably straightforward and avoids the evaluation of (two-loop) Feynman diagrams. It would be interesting to investigate if this relative simplicity persists at higher orders and/or for other classes of theories.

Our second question is `precisely what is allowed in the counterterm action?'. As usual, this is intricately related to the nature of the cutoff and the symmetries that it preserves. Implementing a TCSA cutoff is a non-local operation, and correspondingly the counterterm action could feature non-local terms as well \cite{PhysRevD.91.025005}. Clearly, an arbitrarily non-local counterterm action could be disastrous for the viability of the Hamiltonian truncation method, but fortunately the non-localities are suppressed by powers of the cutoff. In \cref{sec:TCSAdivs}, we use crossing symmetry to analyse the structure of the leading-order divergence. At subleading orders we cannot use any general theorems, but for the $\phi^4$ theory we can make progress by analysing a particular summand; this we do in detail in \cref{sec:scalarthy}. This allows us to demonstrate the necessity of non-local counterterms at second order, as well as tensorial counterterms that in principle could break Lorentz invariance.

Lastly, with an eye towards numerical work we consider the perturbative determination of the coefficients of the counterterm action for the $\phi^4$ theory at second order. When a counterterm is marginally relevant, this may be of limited relevance for practical numerical computations because in such cases the counterterms receive corrections at all orders in perturbation theory and numerical tuning will be required to obtain finite answers in the large cutoff limit. However, when the counterterms are strictly relevant, they receive only a finite number of pertubative contributions and the determination of these coefficients is directly useful for numerical studies. We provide precise expressions in \cref{sec:3d}, including non-divergent subleading terms which can be used to improve the Hamiltonian in numerical studies.


\section{Anomalous Dimensions from Infinite Matrix Diagonalisation}\label{sec:hamvslag}
In this section we explain how the perturbative computation of the anomalous dimensions of composite operators is related to the diagonalisation of the infinite, tree-level matrix of operator product expansion (OPE) coefficients. We start with a CFT in $d$ dimensions and deform it by a relevant operator that we call $\sigma(x)$, with dimension $\Delta_\sigma$. We work perturbatively in the corresponding coupling constant $g$.

\subsection{The Hamiltonian Perspective}
The Hamiltonian procedure starts by putting the $d$-dimensional UV CFT on the cylinder $\mathbb R \times S^{d-1}_R$ where, by virtue of the state-operator correspondence, the Hamiltonian is simply the dilatation operator:
\be
H_{\text{CFT}} \ket{\op_i} = \frac{\Delta_i}{R} \ket{\op_i}\,.
\ee
where $R$ is the cylinder radius and we label states by their corresponding local operator $\op_i(x)$. A relevant deformation of the CFT by an operator $\s$ with coupling $g$ modifies the Hamiltonian to
\be
H = H_{\text{CFT}} + R^{d-1}\int_{S^{d-1}} d {\bf n} \, g\, \s(\t,{\bf n}) + H_\text{ct},
\ee
where $\bf n$ is a unit vector in $\mathbb R^d$ which parametrises $S^{d-1}$. We will be consistent in writing operators in the cylinder picture with two arguments (a `time' component and a unit vector in $\mathbb R^d$) whereas flat space operators will be given one argument (a vector in $\mathbb R^d$). $H_\text{ct}$ is the counterterm Hamiltonian, which we assume starts at order $g^2$.

We can compute the matrix elements of the Hamiltonian in the UV basis by transforming to flat space and using the standard CFT OPE, which takes the form
\be\label{sigmaOPE}
\sigma(x) \op_j(0) = \sum_k \frac{C\du{\sigma j}{k}}{|x|^{\D_\sigma + \D_{jk}}} \op_k(0) + \ldots
\ee
with the dots here representing non-scalar operators (and so k may or may not be primary). Also, we define $\D_{ij}\colonequals \D_i-\D_j$. Then, in terms of
\be\label{OPEMatrix}
V\du{i}{j} \colonequals S_d R^{d- \Delta_\s} g C\du{\s i}{j}
\ee
we find that
\be
H \ket{\op_i} = \frac{1}{R} \left( \Delta_i \delta_i^j + V\du{i}{j} + W\du{i}{j} \right) \ket{\op_j}
\ee
where $R^{-1} W\du{i}{j}$ are the matrix elements of $H_\text{ct}$ in the CFT basis. Here, $S_d\colonequals 2\pi^{d/2}/\G(d/2)$ is the volume of the unit radius sphere embedded in $d$ dimensions.

In order to find the spectum of the deformed theory, we need to diagonalise the Hamiltonian matrix. To second order in the coupling $g$, we find the eigenvalues
\be
\label{Hamevs}
E_i = \frac{1}{R} \left( \Delta_i + V\du{i}{i} + \sum_{j \neq i} \frac{V\du{i}{j} V\du{j}{i}}{\D_{ij}}  + W\du{i}{i} + \ldots \right)
\ee
where the index $i$ is not summed over and we only keep the second-order term in $W\du{i}{i}$. For this equation to be valid we need to request that $V\du{i}{j}$ is diagonal in the finite-dimensional subspace of operators with the same $\Delta_i$, and that any degeneracies are broken by the second-order correction. If these conditions are not met, we need to resort to the usual methods of degenerate perturbation theory to find the correct eigenvalues.

Since the energies are supposed to be finite, the role of the counterterms at second order is to make finite the expression
\be
\sum_{k \neq i} \frac{V\du{i}{k} V\du{k}{j}}{\D_{ik}} + W\du{i}{j} 
\ee
for $i = j$. Notice that it is not necessary for the counterterms to also make this expression finite when $i \neq j$ -- we will comment further on this below. For later reference, we note that the third-order corrections also take a well-known form and are given by
\be\label{3rdOrderEnergy}
\frac{1}{R} \left( \sum_{k \neq i} \sum_{j \neq i} \frac{V\du{i}{j}V\du{j}{k}V\du{k}{i}}{\D_{ij}\D_{ik}}- \sum_{j \neq i} \frac{V\du{i}{i}V\du{i}{j}V\du{j}{i}}{\D_{ij}^2} + \sum_{k \neq i} \frac{W\du{i}{k}V\du{k}{i}}{\D_{ik}} + \sum_{j \neq i} \frac{V\du{i}{j}W\du{j}{i}}{\D_{ij}}+ W\du{i}{i}\right)
\ee
where each instance of the counterterm Hamiltonian $W$ is taken at the appropriate order. 

\subsection{The Lagrangian Perspective}
In the Lagrangian approach, or more precisely in conformal perturbation theory, we compute the perturbative renormalisation of $\op_i$ by introducing the renormalised operator
\be
[\op]_i(x) \colonequals Z\du{i}{j} \op_j(x)
\ee
and perturbatively evaluating correlation functions of the form
\be
\GG_i \colonequals \vev{\ldots [\op]_i(0)}_g \colonequals \vev{\ldots \exp\left( - \int d^d x  \left(g \, \s(x) + L_\text{ct}(x) \right)\right) Z\du{i}{j} \op_j(0)}
\ee
now evaluated on flat $R^d$. Here the ellipses signify a string of operators inserted away from the origin and the expectation values on the right-hand side are those of the undeformed theory. Our aim is to compute the matrix of anomalous dimensions
\be
\Gamma\du{i}{j} \colonequals - \left. \mu \frac{\partial}{\partial \mu} \log Z\du{i}{j}\right|_{g_B}
\ee
where the partial derivative is taken with the bare (dimensionful) coupling $g_B$ held fixed. We assume that the counterterms start at $O(g^2)$ and we work to second order in $g$. We shall also assume that the dimension $\Delta_\sigma$ of the perturbing operator is nearly marginal,
\be
\Delta_\s = d - \e
\ee
and that $\epsilon$ is small, which essentially amounts to working in dimensional regularisation. As usual, we will ignore power law divergences and focus on the poles in $\epsilon$.

The expansions
\be
\begin{split}
	Z\du{i}{j} &= \d_i^j + g Z_{i}^{(1)j} + g^2 Z_{i}^{(2)j} + \ldots \\
	\GG_i &= \GG^{(0)}_i + g\,\GG^{(1)}_i + g^2\,\GG^{(2)}_i+\ldots \\
	L_\text{ct} &= g^2 L_\text{ct}^{(2)}+\ldots
\end{split}
\ee
then give
\be\label{Gdefinition}
\begin{split}
	\GG^{(0)}_i &= \vev{\ldots \op_i(0)}\\
	\GG^{(1)}_i &= \vev{\ldots \left( - \int d^d x \, \s(x)\d_i^j + Z_{i}^{(1)j}\right) \op_j(0)}\\
	\GG^{(2)}_i &=
	\vev{\ldots \Big( \hf \int d^d x \int d^d y \ \s(x) \s(y) \d_i^j - \int d^d x \, L_\text{ct}^{(2)}(x)\d_i^j \\
		& \qquad \qquad \qquad - \int d^d x \,\s(x)Z_{i}^{(1)j} + Z_{i}^{(2)j} \Big) \op_j(0)}
\end{split}
\ee
In calculating $Z\du{i}{j}$ we may focus on some on some neighbourhood of $\op_i$, where the pertinent divergences appear. Therefore we limit all spatial integrals to a spherical region of radius $R$, away from the other operator insertions. For definiteness, one may think of this procedure as the perturbative computation of one-point functions on the ball given by $|x| < R$, but in practice such a physical picture is not important for the computations of the renormalisation constants. 

Upon substitution of the OPE (\cref{sigmaOPE} into \cref{Gdefinition}), we find divergences which we can make finite using dimensional regularisation. Collecting the first-order terms, we find that
\be
\begin{split}
&Z_i^{(1)j} - \int_{|x| < R} d^d x \, \frac{C\du{\s i}{j}}{|x|^{\D_\sigma + \D_{ij}}}  = \\
&Z_i^{(1)j} - \frac{S_{d} C\du{\s i}{j} R^{\e - \D_{ij}}}{\e - \D_{ij}}
\end{split}
\ee
should be finite. We see that there are divergences only when $\D_{ij} = O(\e)$. What's more, since $R$ is a scale which is set far away from the operator insertion, locality dictates that $Z\du{i}{j}$ cannot depend on $R$. We therefore need to introduce a renormalisation scale $\mu$. Altogether, we therefore set
\be
\label{Z1ij}
Z_i^{(1)j} =
\begin{cases}
\displaystyle \frac{S_d C\du{\s i}{j} \mu^{- \e +\D_{ij}}}{\e - \D_{ij}} &\qquad \text{if }\D_{ij} = O(\epsilon)\\
0 &\qquad \text{otherwise}
\end{cases}
\ee
Now we need a little discussion about $C\du{\s i}{j}$. As observed in \cite{PhysRevD.91.025005}, there is a clear problem if $\D_{ij} = \e$, since in that case the integral would not be rendered finite by dimensional regularisation. Therefore $C\du{\s i}{j}$ must vanish precisely for these cases, at least for every theory that is made finite by dimensional regularisation. (This was also explicitly shown to be the case in the $\phi^4$ theories in \cite{PhysRevD.91.025005}.) In fact, if $\D_{ij} = \kappa_{ij} \epsilon$ with $\kappa_{ij}$ finite as $d \to 4$, then $C\du{\s i}{j} = 0$ unless $\k_{ij} = 0$. Furthermore, since operators are orthogonal unless $\D_{ij} = 0$ for all $d$,\footnote{To clarify: here we use the fact that we can track operators $\op_i$ in the free theory whilst varying $d$, so their dimensions $\Delta_i$ then become simple functions of $d$. The claimed orthogonality then follows from the $\phi$\emph{-type} selection rule in \cite{PhysRevD.93.125025}.} it follows that we can pick an orthogonal basis in the space of operators where both the tree-level scaling dimensions are diagonalised for all $d$ and also $C\du{\s i}{j}$ is \emph{diagonal} on every finite-dimensional subspace of operators whose scaling dimensions coincide for $d \to 4$. (Of course, outside of this subspace it can have all kinds of off-diagonal terms.) In this basis, we find the simpler structure
\be
Z_i^{(1)j} =
\begin{cases}
\displaystyle \frac{S_d C\du{\s i}{i} \mu^{-\e}}{\e} &\qquad \text{if }\op_i = \op_j\\
0 &\qquad \text{otherwise}
\end{cases}
\ee
and, again in this basis, the leading order anomalous dimensions are then simply
\be
\Gamma\du{i}{i} = g \mu^{-\e} S_d C\du{\s i}{i} + O(g^2)
\ee
where we used that $g_B = g + O(g^2)$. This expression agrees precisely with the Hamiltonian picture discussed previously, which is \cref{Hamevs} to first order in the coupling. This first-order computation can also be found in the textbook \cite{Cardy.ScalingAndRenormalization}.

At second order, we find some new structures. For now, we will assume that the counterterm Lagrangian is given by a simple renormalisation of the coupling,
\be
g^2 L_{\text{ct}}^{(2)} = g^2 \mu^{-\e} S_d X^\sigma \sigma (x)\,
\ee
with the dimensionless coefficient $X^\sigma$ tuned to make the second-order results finite. (In subsequent sections, we will allow for other operators to appear in the counterterm action.) For future reference, we mention that with this counterterm the bare coupling is
\be
g_B(g,\mu) = g + g^2 \mu^{-\e} S_d X^\sigma + O(g^3)\,.
\ee
Using then that $\hf \int d^d x \int d^d y\, \s(x) \s(y) = \int d^d x \int_{|y| < |x|} d^d y\, \s(x) \s(y)$ to make sure the OPE expansion is valid, we find
\be
\begin{split}
\GG^{(2)}_i &= \Big( \frac{C\du{\s i}{k} C\du{\s k}{j} S_d^2 R^{2 \e -\D_{ij}}}{(\e -\D_{ik})(2 \e-\D_{ij})}
- \frac{Z^{(1)k}_i C\du{\s k}{j} S_d R^{\e - \D_{kj}}}{\e - \D_{kj}} \\
& \qquad \qquad + Z^{(2)j}_i -  \frac{X^\sigma C\du{\s i}{j} S_d^2 \mu^{-\e} R^{\e- \D_{ij}}}{(\e - \D_{ij})} \Big) \vev{\ldots \op_j(0)} 
\end{split}
\ee
and therefore the term in parentheses should be finite. We see that possible divergences can arise through the sum over intermediate operators $k$, but also through small denominators, for example when $\D_{ij}= O(\e)$. The latter divergences are cancelled by setting
\be
\begin{split}
&Z^{(2)j}_i = \\ 
&\begin{cases}
- \frac{C\du{\s i}{k} C\du{\s k}{j} S_d^2 \mu^{- 2 \e +\D_{ij}}}{(\e - \D_{ik})(2 \e-\D_{ij})}
+ \frac{Z^{(1)k}_i C\du{\s k}{j} S_d \mu^{- \e + \D_{kj}}}{\e - \D_{kj}} +  \frac{X^\sigma C\du{\s i}{j} S_d^2 \mu^{-2 \e +\D_{ij}}}{(\e -\D_{ij})} &\qquad \text{if }\D_{ij} = O(\epsilon)\\
0 &\qquad \text{otherwise}
\end{cases}
\end{split}
\ee
This cancellation is not entirely obvious since there are double poles, but upon substituting the lower-order result from \cref{Z1ij}, one finds that all of the divergences are indeed removed. Let us now choose the aforementioned basis of operators, where we find the simpler expression:
\be
\begin{split}
&Z^{(2)j}_i = \\ 
&\begin{cases}
\frac{S_d^2 \mu^{-2\e}}{\e} \left(
- \frac{C\du{\s i}{k} C\du{\s k}{i}}{2 (\e -\D_{ik})}
+ \frac{C\du{\s i}{i} C\du{\s i}{i}}{\e} +  X^\sigma C\du{\s i}{i}\right) &\qquad \text{if }\op_i = \op_j\\
0 &\qquad \text{otherwise}
\end{cases}
\end{split}
\ee
Using this expression to compute the matrix of anomalous dimensions, we find that the double poles cancel precisely and that
\be
\label{Giisecond}
\G\du{i}{i} = V\du{i}{i} + \sum_{k \neq i} \frac{V\du{i}{k}V\du{k}{i}}{\D_{ik} - \e} + W\du{i}{i}  + O(g^3)
\ee
where now
\be
V\du{i}{j} = S_d \mu^{-\e} gC\du{\s i}{j}\,, \qquad \qquad W\du{i}{i} = S_d^2 \mu^{-2\e} X^\s C\du{\s i}{i}\,.
\ee
At an IR fixed point the $\G\du{i}{i}$ will be the anomalous dimensions of the composite operators, so the coefficient $X^\s$ should be chosen such that these are finite for all $i$. That this can be done at all is of course a consequence of perturbative renormalisability.

The third-order correction can be found in the same manner. We will not spell out the details of the tedious but straightforward computation and instead quote the result:
\be \label{3rdorderandim}
\left( \sum_{k \neq i, m \neq i} \frac{V\du{i}{k} V\du{k}{m} V\du{m}{i}}{(\D_{ik} - \e)(\D_{im} - 2\e)} - V\du{i}{i} \sum_{k \neq i}\frac{V\du{i}{k}V\du{k}{i}}{(\D_{ik} -\e)(\D_{ik} - 2\e)} + \sum_{k \neq i} \frac{(W\du{i}{k} V\du{k}{i} + V\du{i}{k} W\du{k}{i})}{\D_{ik} - \e} + W\du{i}{i}\right)
\ee
where we work in the basis discussed above and the third-order counterterm action is assumed to take the form
\be
g^2 L_{\text{ct}}^{(3)} = g^3 \mu^{-2\e} S_d^2 Y^\sigma \sigma (x)
\ee
for some (divergent) c-number $Y^\sigma$.

\paragraph{}

From the renormalisation of composite operators, we can work out the beta function to one higher order using a familiar trick. Consider
\be
\vev{\ldots \exp\left( - g_B(g,\mu) \int d^d x \, \sigma(x) \right)}
\ee
away from any operator insertions, and with the bare coupling $g_B(g,\mu) = g + g^2 \mu^{-\e} S_d X^\s + g^3 \mu^{-2\e} S_d^2 Y^\s + O(g^4)$, which ensures that the result is finite. Taking a derivative with respect to $g$, we find an extra insertion of $\sigma$. This is still finite, so we conclude that
\be
\vev{\ldots \exp\left( - g_B(g) \int d^d x \, \sigma(x) \right) \frac{\partial g_B}{\partial g}  \sigma(0)}
\ee
is also finite. But then we can choose
\be
Z\du{\s}{\s} = \frac{\partial g_B}{\partial g}\,.
\ee
for the renormalisation of the operator $\sigma(x)$. Using that
\be
\beta(g) \colonequals \left. \mu\frac{\partial g}{\partial \mu}\right|_{g_B} 
\ee
and doing a little rewriting, we find the familiar relation between the anomalous dimension and the derivative of the beta function:
\be
\frac{\partial \beta}{\partial g} = \Gamma\du{\s}{\s}
\ee
Below, we will use this to find the beta function at order $g^3$ from the renormalisation factor $Z\du{\s}{\s}$ at order $g^2$.

\subsection{Comparing the Hamiltonian and Lagrangian Approaches}
We have now an abstract way to compute two sets of observables to third order in the deforming operator $g$: the spectrum of the theory on the cylinder expressed in \cref{Hamevs} and \cref{3rdOrderEnergy}, and the matrix of anomalous dimensions in \cref{Giisecond} and \cref{3rdorderandim}. Both expressions are similar and become equivalent if we ignore the additional $\epsilon$ in the denominators and identify $R$ and $\mu^{-1}$.\footnote{Ignoring the $\epsilon$ factors in the denominators is not obviously allowed, since the whole sum is divergent and, after defining it through analytic continuation, has a pole at $\epsilon = 0$. Nevertheless, we find in the next subsection that the finite part is unmodified by the presence of the additional $\epsilon$ in the denominator. Similar cancellations are presumably required for the Hamiltonian and Lagrangian perspectives to agree also at higher orders, but we have not investigated this in detail.}

Altogether, we can view the perturbative computation of anomalous dimensions in a new light: not as the diagonalisation of the finite-dimensional matrices $\G\du{i}{j}$ whose elements we need to compute order by order in perturbation theory, but rather as the perturbative diagonalisation of the infinite-dimensional matrix $V\du{i}{j}$ whose elements are just those of the unperturbed theory.

We should stress that the two pictures do not always have to agree. In fact, they agree if two conditions are met. First of all, the perturbing operator should be marginal or marginally (ir)relevant. Indeed, if this condition is not met there are (generically) no logarithmic divergences, there is no renormalisation scale $\mu$, and the matrix of anomalous dimensions vanishes. In our computations, this shows up because we need poles in $\epsilon$ in the Lagrangian computation but not in the Hamiltonian one. Secondly, the theory should remain conformal or flow to a nearby IR (or UV) fixed point. Only in this case can we use the Callan-Symanzik equation to relate $\G\du{i}{j}$ to scheme-independent observables like the anomalous dimensions of local operators, and of course the state-operator map to relate these scaling dimension to the spectrum of the theory on the cylinder.

Both of the conditions discussed above are met in Wilson-Fisher type fixed points, to which we will turn our attention in \cref{sec:scalarthy}. This will also allow us to investigate the counterterm action $W\du{i}{j}$ in more detail. To do so we however first need to improve our understanding of the generally infinite sum in \cref{Hamevs}, which we will discuss in the next section.


\section{Divergences with the TCSA cutoff} \label{sec:TCSAdivs}
In a nutshell, the TCSA procedure of Yurov and Zamolodchikov \cite{Yurov:1989yu} amounts to truncating the Hamiltonian matrix by ignoring all states (in the UV basis) with dimensions larger than a cutoff value $\Delta_{\max}$. The resulting matrix can then be diagonalised numerically, which for sufficiently small couplings (made dimensionless by using powers of $R$) ought to give an accurate representation of the spectrum of the deformed theory.

On the cylinder ${\mathbb R} \times S^{d-1}$ this truncation procedure preserves the rotations $SO(d)$ and time translations $\mathbb R$, so in principle the regularisation prescription does not break more symmetries than the background geometry, which itself serves as an infrared regulator. The counterterms that we find will therefore preserve these symmetries, but a priori one may not recover a full Lorentz symmetry as we send the sphere radius to infinity. Another issue is that the truncation of the Hilbert space breaks locality on the sphere, so locality of the counterterm action is no longer guaranteed. The two issues of non-Lorentz invariant counterterms and non-local counterterms were raised before in \cite{PhysRevD.91.025005}.

In the radial quantisation picture, we can mimic the cutoff in the TCSA by sandwiching the insertions of the perturbing operator between a Hilbert space projector
\be
\mathcal{P}\colonequals\sum_{\D_n\leq\D_{\max}}\ket{\op_n}\bra{\op_n}
\ee
which removes intermediate states of weight greater than $\D_{\max}$. Notice that this amounts to a cutoff in energies with associated scale $\Lambda = \D_{\max} / R$, with $R$ the radius of the sphere where we insert $\mathcal P$. This cutoff breaks locality on these spatial spheres (but not in the radial `time' direction), as well as covariance under translations on the plane.

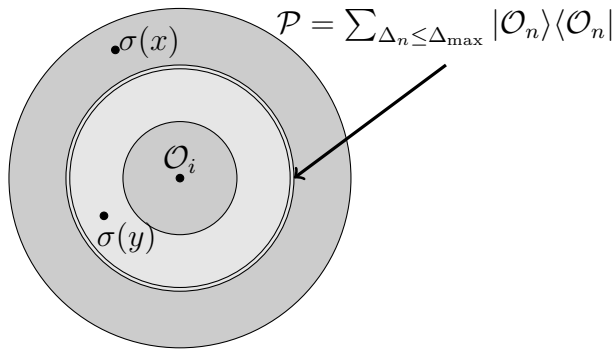
\begin{figure}
	\begin{center}
		\begin{tikzpicture}
		\filldraw [fill=gray!40,draw=black] (0,0) circle (2.25);
		\filldraw[fill=gray!20,draw=black] (0,0) circle (1.5);
		\filldraw[fill=gray!20,draw=black] (0,0) circle (1.45);
		\filldraw [fill=gray!40,draw=black] (0,0) circle (0.75);
		\filldraw[fill=black,draw=black] (0,0) circle (0.05);
		\filldraw[fill=black,draw=black] (-1,-0.5) circle (0.05);
		\filldraw[fill=black,draw=black] (-0.85,1.7) circle (0.05);
		\draw [<-,very thick] (1.5,0)--(3.5,1.5);
		\node at (0,0.25) {$\op_i$};
		\node at (-0.7,-0.8)  {$\s(y)$};
		\node at (-0.4,1.8) {$\s(x)$};
		\node at (3.5,2) {$\mathcal{P}=\sum_{\D_n\leq\D_{\text{max}}}\ket{\op_n}\bra{\op_n}$};
		\end{tikzpicture}
	\end{center}
	\caption{A Hilbert space projector $\mathcal{P}$ is placed between the two $\s$ operators, removing intermediate states of weight greater than $\D_{\text{max}}$ and regulating contact divergences.}
	\label{projectorpic}
\end{figure}

To take the new cutoff into account we need to modify the analysis in the preceding section as follows. First of all, we need to more carefully keep track of power law divergences and not just the poles as $\e\to 0$ in dimensional regularisation. This implies that, compared to \cref{sec:hamvslag}, we need to modify the counterterm action to
\be
g^2 L_\text{ct}^{(2)}(x) = g^2 \m^{-\e}S_d X^k\op_k(x) + g^2 \m^{-\e} S_d X^\s \s(x)
\ee
with some operators $\op_k(x)$ and coefficients $X^k$ and $X^\s$. We expect $L_\text{ct}$ to include $\D_{\max}$ divergent counterterms defined to cancel divergences arising from the hard truncation of the Hilbert space, as well as $1/\e$ divergent counterterms associated with the same coupling renormalisation. In terms of the double limit, we should technically take $\D_{\text{max}}\rightarrow\infty$ with $\e$ finite prior to taking $\e\rightarrow0$. We will however only analyse the $\Delta_{\max}$ divergences, which suffices to get a finite answer at fixed $\epsilon$, so e.g. at $\epsilon = 1$ for the $\phi^4$ theory in $d = 3$.\footnote{Notice that we should also be able to renormalize the second-order divergences for $\epsilon < 0$ -- there is a suitable counterterm action for the $\phi^4$ theory for any $d$ to any finite order in perturbation theory.}

The coefficients $X^k$ and $X^\s$ are non-trivial functions of the cutoff, and are fixed by the requirement that physical observables are finite. In the Hamiltonian perspective (to second order) this concretely means that there should be no divergences in
\be \label{divsum}
\sum_{\substack{k \neq i \\ \D_k \leq \D_{\max}}} \frac{C\du{i}{k}C\du{k}{i}}{\D_{ik}} + X^k C\du{ki}{i} + X^\s C\du{\s i}{i}
\ee
as we send $\D_{\max} \to \infty$. Clearly, in order to determine the counterterms, we need to have some amount of control over the asymptotics of the sum. We will focus on the behaviour of the summand as a function of the dimension of the intermediate operator, that is we will consider the object:
\be
\sum_{k: \Delta_k = \Delta} C\du{\s i}{k} C\du{\s k}{j}
\ee
as a function of $\Delta$ (and $i$ and $j$).

In this section, we will review the analysis in \cite{PhysRevD.91.025005}, which uses crossing symmetry and a Tauberian theorem to constrain the large $\Delta$ behavior in full generality. In \cref{subsec:scalarTCSA}, we will then apply the results to the free scalar theory, and show that there we can obtain better results than those rigorously proven by the Tauberian theorem.

\subsection{The Tauberian Theorem}\label{subseccrossing}
In an attempt to estimate the divergences in \cref{divsum}, we introduce the four-point function studied in \cite{PhysRevD.91.025005}
\begin{equation}\label{4point}
\mathcal{F}_{ji}(\t) \colonequals e^{\t(\D_\s+ \D_{ij}/2)} \vev{\op_j(\infty)\int_{S^{d-1}} d{\bf n}\int_{S^{d-1}} d{\bf n}' \,\s(e^{\t/2}{\bf n})\s(e^{-\t/2}{\bf n}')\op_i(0)}
\end{equation}
where $\t>0$ and $\op(\infty)\colonequals \lim_{|x|\rightarrow\infty}|x|^{2\D_\op}\op(x)$. The exponential pre-factor is pulled out for later convenience. Evaluating this gives
\begin{equation}\label{schan}
\mathcal{F}_{ji}(\t) = S_d^2  e^{\t(d-\e+\D_i)}\sum_{k} e^{-\tau \D_k}C\du{\s i}{k}C\du{\s k j}{}
\end{equation}
with $C\du{\s k j}{} = \sum_l C\du{\s k}{l} G_{lj}$, where $G_{l j}$ is the Gram matrix (which will drop out from all our relevant results below). As in \cref{sec:hamvslag}, the sum is over intermediate scalars only, because of the spherical integrals. We can try to get an idea of the asymptotic behavior of the sum by using an inverse Laplace transform. For example, if the behaviour near $\t = 0$ is of the form
\begin{equation}
\mathcal{F}_{ji}(\t) = c_\a \tau^{-\a} (1 + O(\tau))
\end{equation}
with $\a > 0$, then we would roughly speaking expect that
\begin{equation}\label{asymptsumnaive}
\sum_{k \,:\, \Delta_k = \Delta} C\du{\s i}{k} C\du{\s k j} \sim \,\frac{c_\a \Delta^{\a -1}}{\G(\a)} \qquad \text{as }\D \to \infty,
\end{equation}
simply because
\begin{equation}
\int_0^\infty \Delta ^{\alpha -1} e^{-\D \t} d\Delta = \Gamma (\alpha ) \tau^{-\a}
\end{equation}
and the non-analytic behavior in $\tau$ originates from the large $\D$ part of the integral.

Of course, the preceding claim cannot be exactly true, for the simple reason that the left-hand side of \cref{asymptsumnaive} is not a smooth function of $\D$. The precise statement follows from the Hardy-Littlewood Tauberian theorem, which states that this holds only in an aggregrate sense. The version that we will need is the one explained in \cite{Pappadopulo:2012jk} and proven, for example, in \cite{feller1971introduction}: take a (positive) measure $d \mu (\D)$ such that $\int_a^b d\mu(\D)$ is finite for every finite $a$ and $b$. Now if
\begin{equation}
F(\t) = \int_0^\infty e^{-\t \D} d\mu(\D)
\end{equation}
behaves for small $\t$ as 
\begin{equation}
F(\t) \sim \t^{-\rho}
\end{equation}
with $\rho > 0$, then 
\begin{equation}
\int^{\D_{\max}} d\mu(\D) \sim \frac{\D_{\max}^\rho}{\G(\rho + 1)}
\end{equation}
for large $\D_{\max}$. Here $a \sim b$ means that $a/b \to 1$ in the relevant limit.  

Unfortunately, without further assumptions we can say little useful about the subleading terms. For example, if we try to subtract the leading term from $d\mu(\D)$, then it is generally no longer positive and the theorem ceases to apply.

\subsection{Using Crossing Symmetry}\label{subsec:UsingCrossing}
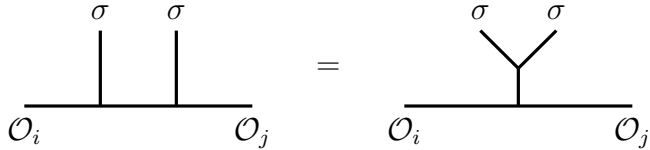
\begin{figure}
	\begin{center}
		\begin{tikzpicture}
		\draw[very thick] (0,0) node [align=left, below] {$\op_i$} -- (3,0) node [align=right, below] {$\op_j$};
		\draw[very thick] (1,0) -- (1,1) node [align=center, above] {$\s$};
		\draw[very thick] (2,0) -- (2,1) node [align=center, above] {$\s$};
		\draw (4,1/2) node {$=$};
		\draw[very thick] (5,0) node [align=left, below] {$\op_i$} -- (8,0) node [align=right, below] {$\op_j$};
		\draw[very thick] (13/2,1/2) -- (6,1) node [align=center, above] {$\s$};
		\draw[very thick] (13/2,1/2) -- (7,1) node [align=center, above] {$\s$};
		\draw[very thick] (13/2,0) -- (13/2,1/2);
		\end{tikzpicture}
	\end{center}
	\caption{Crossing symmetry allows us to equate fusion channels. This is integral in the arguments of \cref{sec:TCSAdivs} and \cref{sec:scalarthy} which explain why counterterms are made up from the local operators in the $\s\times\s$ OPE, dressed with dilatation operators.}
	\label{fig:crossingpic}
\end{figure}

To estimate the small $\tau$ behaviour of \cref{4point}, we expand in the crossed channel by first fusing the two $\s$ operators, as indicated in \cref{fig:crossingpic}. We then obtain
\begin{align} \label{tchan}
\mathcal{F}_{ji}(\t) &= \int_{S^{d-1}} d{\bf n}\int_{S^{d-1}} d{\bf n}'  \sum_{\hat k} \frac{C\du{\s \s}{\hat k} C_{\hat k i j}}{( {\bf n} -e^{-\t} {\bf n}')^{2 \D_\s}} G_{\D_{\hat k}}^{(\ell_{\hat k})} \left(\D_{ij}; u , v\right)\\
&= S_{d} S_{d-1}\sum_{\hat k} \int_0^\pi d\theta \frac{\sin(\theta)^{d-2} C\du{\s \s}{\hat k} C_{\hat k i j}}{(1+e^{-2\t}-2e^{-\t}\cos \q)^{d-\e}} G_{\D_{\hat k}}^{(\ell_{\hat k})} \left(\D_{ij} ; u,v\right)
\end{align}
with $\cos(\theta) = {\bf n} \cdot {\bf n}'$ and with
\begin{equation}
u = ({\bf n} - e^{-\t}{\bf n}')^2 = 1+e^{-2\t}-2e^{-\t}\cos \q \qquad \qquad v = e^{-2 \t}\,.
\end{equation}
This time, the sum is over all of the \emph{primaries} in the theory, which we have indicated with a hatted index $\hat k$. The conformal blocks are fixed by conformal symmetry and encode the contribution of the descendants. For the decomposition, we set $x_1 = e^{\t/2} {\bf n}$, $x_2 = e^{-\t/2}{\bf n}'$, $x_3 = 0$ and $x_4 = \infty$ and we followed the familiar conventions of Dolan and Osborn \cite{Dolan:2000ut,Dolan:2003hv}. This OPE expansion is not strictly valid across the whole integration domain because the operator at the origin sits midway between the two $\s$ operators when they are at antipodal points. However, we will only be interested in the non-analytic part as $\tau\rightarrow0$, which comes from the $\sigma \time \sigma$ OPE region where $\cos(\theta)$ is close to 1, and in this region the sum converges.

A conformal block can be expanded \cite{PhysRevD.87.106004} as a sum of Gegenbauer polynomials:
\begin{equation}
G_\D^{(l)}(\D_{ij}; z,\bar z)=|z|^\D\sum_{n,m=0}^{\infty}c_{n,m}|z|^n \frac{m!}{(d-2)_m}C_m^{d/2-1}(\cos(\text{arg}(z)))
\end{equation}
The coefficients $c_{n,m}$ depend on $\D_{ij},l,d$ and $\D$. In \cref{CBRR}, we review how to determine $c_{n,m}$ recursively from the Casimir equation.

The complex cross-ratio in our conventions is given by 
\begin{equation}
z = 1 - e^{-\tau + i \theta}
\end{equation}
and therefore
\begin{equation}
|z|^2 = u, \qquad \cos(\arg z) = \frac{1+u-v}{2\sqrt{u}}=\frac{1-e^{-\t}\cos \q}{\sqrt{1+e^{-2\t}-2e^{-\t}\cos \q}}
\end{equation}
This leads us to define the integrals:
\begin{equation}
I^{(m)}_{\alpha} (\tau) := S_d S_{d-1} \int_0^\pi d\theta \frac{\sin(\theta)^{d-2}}{(1+e^{-2\t}-2e^{-\t}\cos \q)^{d-\e-\a/2}} C_m^{d/2-1}(\cos \arg z)
\end{equation}
which have small $\tau$ behaviour of the form, for example,
\begin{align}
I^{(0)}_{\alpha} (\tau) &= \t^{-(d+1-\a-2\e)}S_d^2 \left(\xi(\a)+O(\t)\right)\nonumber\\
I^{(1)}_{\alpha} (\tau) &=  \t^{-(d+1-\a-2\e)}S_d^2 \left(-(d-2)\xi(\a-1)+O(\t)\right)\\
I^{(2)}_{\alpha} (\tau) &=  \t^{-(d+1-\a-2\e)}S_d^2 \left(\frac{(d-2)(d-\a-2\e)}{1+d-\a-2\e}\xi(\a-2)+O(\t)\right)\nonumber
\end{align}
where we have introduced 
\begin{equation}
\xi(\a) \colonequals \frac{\G (\frac{d}{2}) \G (\frac{d+1-\a}{2}-\e)}{2 \sqrt{\pi } \G (d-\frac{\a}{2}-\e )}
\end{equation}
The integral of a conformal block is then an infinite sum of these integrals:\footnote{In practice we will expand this expression around $\tau = 0$, which means we restrict ourselves to only the leading terms in this sum. We are therefore not worried about convergence of the sum.}
\be
{\mathcal I}_{\Delta}^{(\ell)}(\Delta_{ij};\tau) \colonequals \sum_{n,m=0}^{\infty}c_{n,m} \frac{m!}{(d-2)_m}I^{(m)}_{\Delta + n}(\tau)
\ee
and now we can efficiently write
\be
\mathcal{F}_{ji}(\t) = \sum_{\hat k} C\du{\s \s}{\hat k} C_{\hat k i j} {\mathcal I}_{\Delta_{\hat k}}^{(\ell_{\hat k})}(\D_{ij};\tau) 
\ee
For concreteness, an integrated scalar block looks like
\begin{align}
{\mathcal I}_{\Delta}^{(0)}(\D_{ij};\tau) &= I_\D^{(0)}(\t)+\frac{\D_{ij}+\Delta}{2(d-2)}I_{\D+1}^{(1)}(\t)\nonumber\\ &\phantom{=}+\frac{(2+\Delta)(\D_{ij}+\Delta)(2+\D_{ij}+\Delta)}{4d(d-2)(1+\Delta)}I_{\D+2}^{(2)}(\t)\nonumber\\
&\phantom{=}-\frac{(d-2-\Delta)(d-2-\Delta-\D_{ij})(\D_{ij}+\Delta)}{4d(d-2(1+\Delta))}I_{\D+2}^{(0)}(\t)+\ldots
\end{align}
which we can subsequently expand for small $\tau$ using the previous expressions.

The leading term in the $\sigma$ self-OPE corresponds to the identity operator. Its entire contribution to $\FF_{ji}(\tau)$ is
\begin{align}
\FF_{ji}(\tau)& \supset C\du{\s\s}{\bf 1} G_{ij} {\mathcal I}_0^{(0)}(\D_{ij}; \tau) = C\du{\s\s}{\bf 1} G_{ij} I_{0}^{(0)}(\t) \nonumber\\
&\qquad = C\du{\s\s}{\bf 1} G_{ij}\t^{-(d+1-2\e)}S_d^2 \xi(0)\left( 1 + O(\t) \right)
\end{align}
If we set $j = i$ to ensure positivity, we find that the Hardy-Littlewood Tauberian theorem rigorously applies and therefore (with no sum over $i$)
\begin{equation}
\label{idcontr}
\sum_{k\,:\,\D_k \leq \D_{\max} + d - \e + \D_i} C\du{\s i}{k}C\du{\s k}{i} \sim \frac{C\du{\s\s}{\bf 1} \Delta_{\max}^{d+1-2\e}}{\Gamma(d+2-2\e)} \left(\xi(0)+ \ldots \right)
\end{equation}
where the offset in the sum on the left-hand side (which is actually subleading here) originates from the shift in the exponent in \cref{schan}.

\cref{idcontr} is as far as rigorous results can carry us \cite{Pappadopulo:2012jk}. In \cref{subsec:scalarTCSA} we will however show that we can do much better by following \cite{PhysRevD.91.025005} and inverse Laplace transforming the subleading terms in $\tau$ to estimate the subleading terms in the $\Delta_{\max}$ expansion. Below, we use the closed-form OPE coefficients for the $\phi^k$ operators in the free scalar theory to illustrate this procedure in detail.


\section{The Scalar Theory}\label{sec:scalarthy}
In this section, we will exemplify the abstract computations of \cref{sec:hamvslag} and \cref{sec:TCSAdivs} by considering the second-order corrections in a theory of an interacting scalar field $\phi(x)$ in $d$ dimensions. We perturb the free massless theory by the operator $\phi^k(x)$ and investigate the anomalous dimensions of the subset of operators $\phi^l(x)$ with $l \in \{1,2,3,\ldots\}$. The scalar is normalised such that
\be
\vev{\phi(x) \phi(0)}  = \frac{1}{|x|^{2 \D_\phi}}
\ee
in the unperturbed theory. Here $\Delta_\phi = (d-2)/2$. 

We now compute, using Wick contractions\footnote{We use colons to explicitly mark normal ordered operators only when there is a potential ambiguity}, the following OPE:
\be
\begin{split}
&\int_{S^{d-1}} d\textbf{n} \, \phi^k(x)\phi^l(0) = \sum_p p! \binom{k}{p} \binom{l}{p} \int_{S^{d-1}} d\textbf{n}\, |x|^{-2 p \Delta_\phi}:\phi^{k-p}(x)\phi^{l-p}(0): \\
&\qquad= S_d \sum_p p! \binom{k}{p} \binom{l}{p} \sum_{n=0}^\infty \frac{|x|^{2(n- p \D_\phi)}}{2^{2n}n!(d/2)_n}\phi^{l-p} \square^n \phi^{k-p}(0)
\end{split}
\ee
where the integral serves to project onto scalar operators and $(a)_n:=\G(a+n)/\G(a)$ is the usual Pochhammer symbol. The OPE coefficients can then be read off:
\be
\label{opecoeffs1}
C\du{\phi^k \,\, \phi^l}{\phi^{l-p} \square^n \phi^{k-p}} = \binom{k}{p} \binom{l}{p} \frac{p!}{2^{2n}n!(d/2)_n}
\ee
and a similar computation yields, for $n \geq 0$ and $r=k-p+\frac{l-q}{2}\in \mathbb{Z}$,
\be
\label{opecoeffs2}
C\du{\phi^k \,\, \phi^{l-p} \square^n \phi^{k-p}}{\phi^q} = \sum_m r! \binom{k}{r} \binom{k-p}{m} \binom{l-p}{r-m} 2^{2n} (m \D_\phi)_n ((m-1) \D_\phi)_n 
\ee
Notice in particular that
\be
C\du{\phi^4 \,\, \phi^l}{\phi^l} = 6 l(l-1), \qquad \qquad C\du{\phi^3 \,\,\phi^l}{\phi^l} = 0,
\ee
which we will need below.

We can now compute the corrections to the cylinder energies by summing them as in \cref{Hamevs}, and we therefore would like to compute:
\be \label{Xi}
\Xi\du{k\,l}{q} \colonequals \sum_{n = 0}^{\infty} \sum_{p=0}^{\min(k,l)}  \frac{C\du{\phi^k \,\, \phi^l}{\phi^{l-p} \square^n \phi^{k-p}} C\du{\phi^k \,\, \phi^{l-p} \square^n \phi^{k-p}}{\phi^q} }{l \D_\phi - (l + k - 2 p) \Delta_\phi - 2n}
\ee
It is remarkable that we can find a simple closed-form expression for this sum where every intermediate operator is clearly identifiable. As we will see below, this offers us a unique playground to test the ideas introduced in the previous sections without having to resort to any numerical approximations.

The sum in \cref{Xi} is infinite and we will regularise it in three different ways. Our first regularisation procedure is the familiar dimensional regularisation, which will allow us to check our computations and recover the perturbative anomalous dimensions at the Wilson-Fisher fixed points. 


\subsection{Dimensional Regularisation and the Wilson-Fisher Fixed Points}
For each $p$, the infinite sums in \cref{Xi} turn out to be of a ${}_3 F_2$ hypergeometric nature, and after using some hypergeometric identities we can perform the required analytic continuation in $\epsilon$. Collecting all the factors and the lower-order terms as in \cref{Hamevs}, we find that the second-order energies on the cylinder are given by
\be
\begin{split}\label{CylinderEngergies}
&R E_{\phi^l} = l + 6 l(l-1) g R^{-\e} S_d \\ & + \left(- \frac{216}{\e} l (l-1) -68 l^3+132 l^2-52 l + 6 X^\s l (l-1) + O(\epsilon) \right) g^2 R^{-2\e} S_d^2 + O(g^3).
\end{split}
\ee
In the Lagrangian perspective we are supposed to perform as in \cref{Giisecond} which differs from \cref{Xi} by an additional $\epsilon$ in the denominator of the sum. This happens to make the sum slightly easier since we get just ${}_2 F_1$ hypergeometric sums, and we find that the resulting small $\epsilon$ expansion up to $O(\epsilon)$ is exactly the same, so
\be
\begin{split}
&\Gamma\du{\phi^l}{\phi^l} = 6 l(l-1) g R^{-\e} S_d \\ & + \left(- \frac{216}{\e} l (l-1) -68 l^3+132 l^2-52 l + 6 X^\s l (l-1) + O(\epsilon) \right) g^2 R^{-2\e} S_d^2 + O(g^3)
\end{split}
\ee
This result is in agreement with the above discussion: the Lagrangian and Hamiltonian perspectives match for small $\e$. That is, $R E_{\phi^l}=l+\G\du{\phi^l}{\phi^l}+O(\e)$ up to second order in $g$. 

Finding the counterterm action is straightforward, as $X^\s$ is the only counterterm coefficient that we can tune. Notice that it should remove the divergences for every $l$, but this works out perfectly and all anomalous dimensions (or cylinder energies in \cref{CylinderEngergies}) are finite if we set
\be
X^\s = \frac{36}{\e}\,.
\ee

It is worthwhile to work out the details a bit further and get the two-loop anomalous dimensions. 
For $l = 4$, we find
\be
\G\du{\phi^4}{\phi^4} = 72 S_d g \mu^{-\e} - 2448 S_d^2 g^2 \mu^{-2\e}
\ee
and therefore
\be
Z\du{\phi^4}{\phi^4} = 1+\frac{72}{\epsilon } S_d g \mu ^{-\epsilon } + 72 \left(\frac{36}{\e^2} - \frac{17}{\e}\right) S_d^2 g^2 \mu ^{-2 \epsilon } +O\left(g^3\right)
\ee
which we can integrate once more with respect to $g$ to see that
\be
g_B(g,\mu) = g+\frac{36}{\epsilon } S_d g^2 \mu ^{-\epsilon } + 24 \left(\frac{36}{\e^2} - \frac{17}{\e}\right) S_d^2 g^3 \mu ^{-2 \epsilon } +O\left(g^4\right)
\ee
The leading term already agrees with the counterterm we found above, but we now also have the next-order counterterm at our disposal. Similarly, the quantum beta function for the dimensionful coupling is $\beta(g) = 36 S_d g^2 \mu^{-\e} - 816 S_d^2 g^3 \mu^{-2\e} + O(g^4)$ from integrating $\G\du{\phi^4}{\phi^4}$ and so the dimensionless coupling $\tilde g \colonequals g \mu^{-\e}$ has a beta function of the form
\be
\tilde \beta(\tilde g) = - \epsilon \tilde g + 36 S_d \tilde g^2 - 816 S_d^2 \tilde g^3 + O(\tilde g^4)
\ee
The fixed point is located at
\be
\tilde g^*=\left(\frac{\e}{36}+\frac{17\e^2}{972}\right)/S_d
\ee
resulting in the fixed-point anomalous dimensions:
\be
\G^{*\,\phi^l}_{\phi^l}=\frac{\e l(l-1)}{6}-\frac{\e^2 l(47+l(17l-67))}{324}\,.
\ee
Plugging in $l=1,2,3,4$ then gives the familiar results
\begin{align}
\G^*_\phi&=\e^2/108\quad&\G^*_{\phi^2}&=\e/3+19\e^2/162\nonumber\\
\G^*_{\phi^3}&=\e+\e^2/108\quad&\G_{\phi^4}^{*}&=\b'(g^*)=\e-17\e^2/27
\end{align}
The similarity between $\G_{\phi}^*\text{ and }\G_{\phi^3}^*$ arises due to the fact that the conformal multiplets of $\phi^3$ and $\phi$ recombine at the Wilson-Fisher fixed point, according to the equation of motion.

In exactly the same manner we find the following results for the $\phi^3$ theory in $d = 6 - 2 \e$:
\be
\G\du{\phi^l}{\phi^l} = 3l (6 - 5 l) g^2 S_d^2 \mu^{-2\e}
\ee
Notice that $C\du{\phi^3\,\,\phi^l}{\phi^l} = 0$ and therefore the operators do not `see' the $\phi^3$ counterterm. This implies that the sums in $\G\du{\phi^l}{\phi^l}$ have to come out finite, as indeed they do. We deduce:
\be
\tilde \beta(\tilde g) = - \e \tilde g - 27 S_d^2 g^3 
\ee
and find a (non-unitary) fixed point at $g^* S_d = \sqrt{- \e/27}$, leading to the anomalous dimensions
\be
\G^*_\phi=- \e/9\quad\G^*_{\phi^2}=8\e/9\quad\G^*_{\phi^3}=3\e\quad\G_{\phi^4}^{*} =\b'(g^*)=56 \e/9
\ee
and this time the first two are related precisely such that $\Delta_{\square \phi} = \Delta_{\phi^2}$, as expected by the equation of motion. Notice that we did not use the wave function renormalisation counterterm here.



\subsection{The TCSA cutoff}\label{subsec:scalarTCSA}
We have seen that the sum in \cref{Xi} can be regulated by dimensional regularisation, which allowed us to find a somewhat novel way to extract the correct second-order anomalous dimensions. Our main focus in this paper is however the TCSA cutoff introduced in \cref{sec:TCSAdivs}. In this section we will therefore use this cutoff to regularise the sum in \cref{Xi}.

\subsubsection{Determination of the Counterterm Action}
\begin{figure}
\begin{center}
\includegraphics[width=7cm]{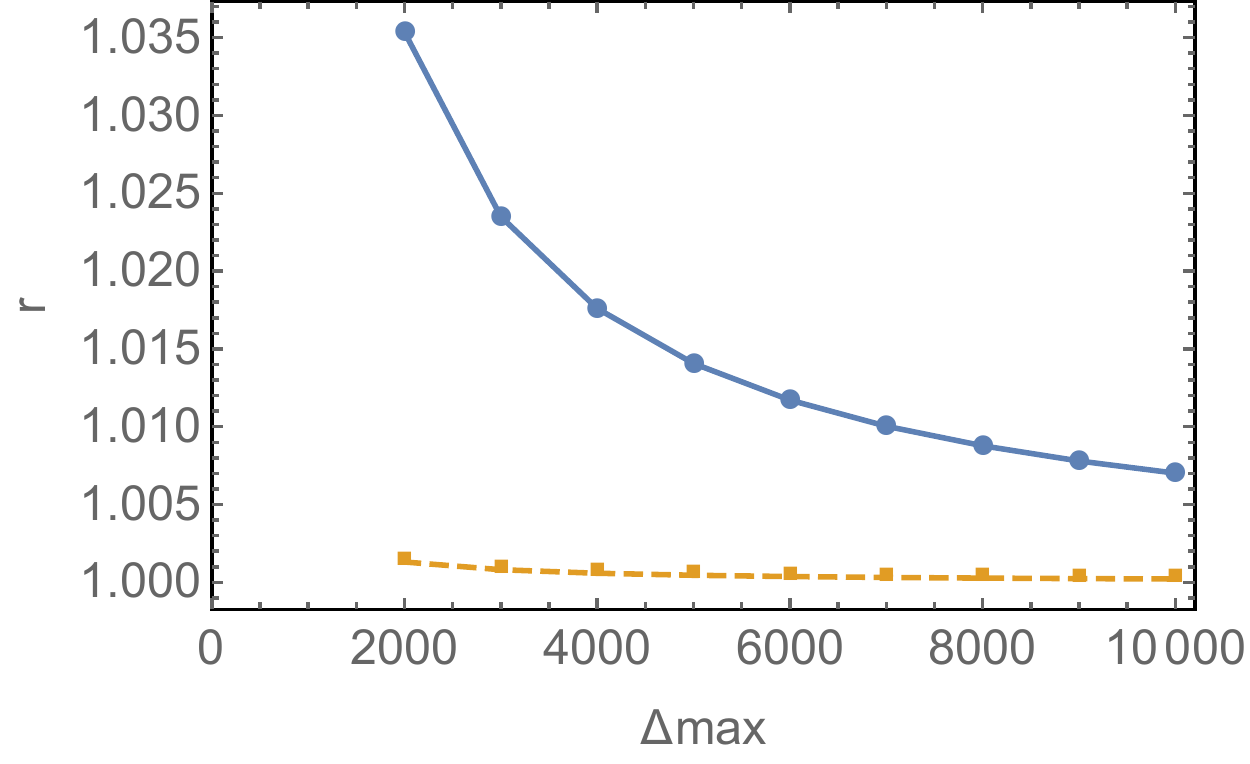}$\quad$
\includegraphics[width=7cm]{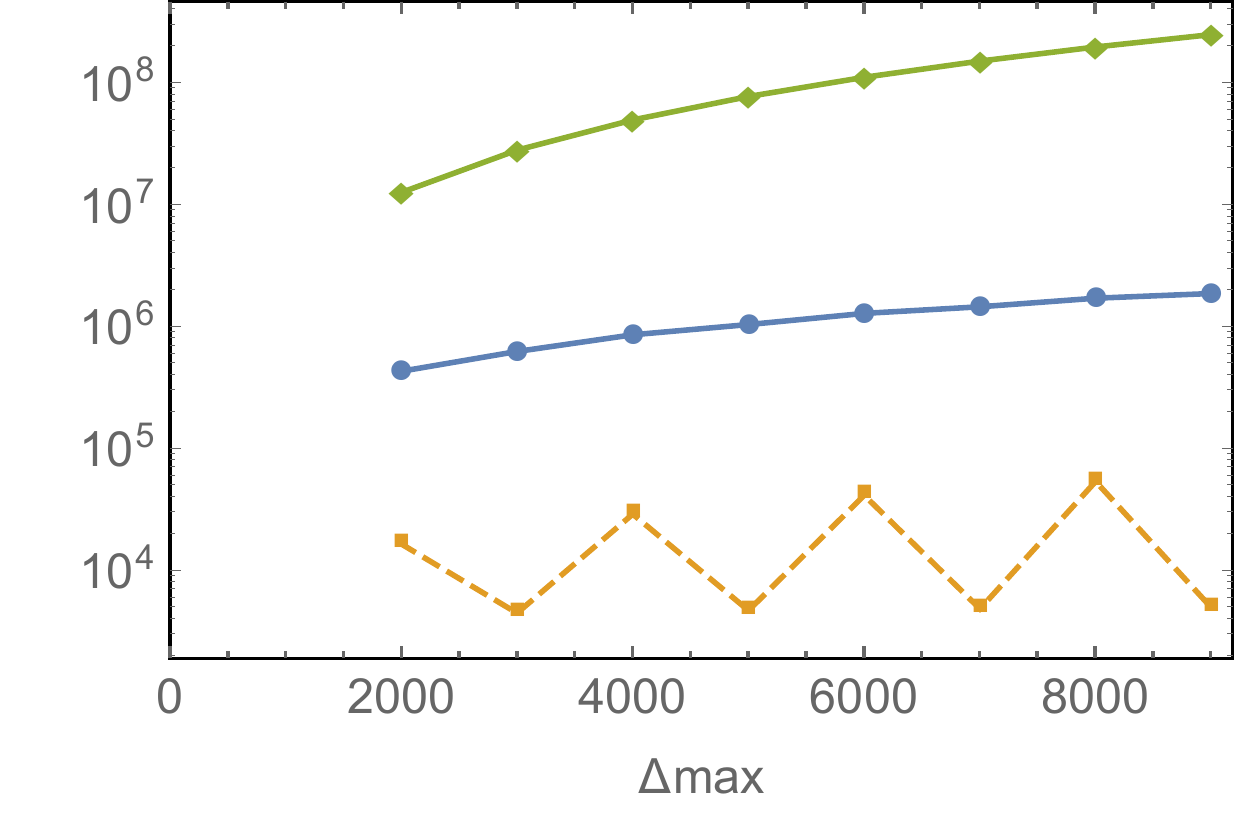}$\quad$
\caption{\label{fig:sum_leading_beh}Tests of the approximation in \cref{sumasympt} with $l = k = 2$ and $\epsilon = 1$ so $d = 3$. As a function of $\D_{\max}$, we plot on the left the ratio $r$ between the left-hand side and the right-hand side (solid line) as well as an improved estimate obtained by including on the right-hand side also the two terms labelled $\#2$ (dashed line). On the right, we log-plotted (from top to bottom) first the value of the left-hand side and then the values obtained by subtracting the terms $\#1$ and the terms $\#1$ and $\#2$, respectively.}
\end{center}
\end{figure}

\cref{subseccrossing} instructs us to consider the four-point function $\FF_{ji}(\tau)$, which is schematically $\vev{j|\int\sigma \int\sigma |i}$. In our case $\sigma(x) = \phi^4 (x)$ and we will take $\op_j$ to be $\phi^k$ and $\op_i$ to be $\phi^l$. Terms in the self-OPE of the $\phi^4$ operator like
\be
\begin{split}
\phi^4(x) \phi^4(0) &\supset \frac{24}{|x|^{8 \D_\phi}} {\mathbf 1} + \frac{96}{|x|^{6\D_\phi}}\left( \phi^2(0) + \text{desc.} \right)  + \frac{72}{|x|^{4\D_\phi}} \left(\phi^4(0) + \text{desc.}\right) + \ldots
\end{split}
\ee
and similarly for the stress tensor $T$, lead to small $\tau$ behavior dictated by the expansion
\be
\FF\ud{j}{i}(\t)\supset 24 \delta^j_i \mathcal I_{0}^{(0)}(\t) + 96 C\du{\phi^2\,i}{j}  \mathcal I_{2 \D_\phi}^{(0)}(\tau) + 72 C\du{\phi^4\,i}{j} \mathcal I_{4\D_{\phi}}^{(0)}(\tau)+ C\du{\phi^4 \phi^4}{T}C\du{T\,i}{j} \mathcal I_d^{(2)}(\tau)+ \ldots
\ee
which concretely leads to small $\tau$ behaviour of the form
\be
\label{FjismalltauWF}
\begin{split}
	S_{d}^{-2} \FF\ud{j}{i}(\t) = 24\delta^j_i \tau^{3 \epsilon - 5} \xi(0)&\left(\underbrace{1}_{\#1}+\underbrace{(4-2\e)\t}_{\#2} +\frac{1}{6}(2-\e)(23-12\e)\t^2+\ldots \right)\\
	+\, 96C\du{\phi^2\,i}{j} \tau^{2 \epsilon - 3} \xi(2-\e)&\left( \underbrace{1}_{\#2} + \left(4-2\e+\frac{\D_{ij}}{2}\right)\t +\text{horrid}\,\t^2+ \ldots \right)\\
	+\, 72C\du{\phi^4\,i}{j} \tau^{\epsilon - 1} \xi(4-2\e)&\left( 1 + \left(4-2\e+\frac{\D_{ij}}{2}\right)\t  + \ldots \right)\\
	+\, C\du{\phi^4 \phi^4}{T}C\du{T\,i}{j} \tau^{2 \epsilon - 1}\xi(4-\e)  &\left(-\frac{2\e}{4-3\e}  +\, \ldots\right)
\end{split}
\ee
where
\begin{equation} \label{horrid}
\text{horrid}=
\frac{3 \Delta_{ij}^2 \epsilon -12 \Delta_{ij} (\epsilon -2) (2 \epsilon -1)+2 (\epsilon -2) (\epsilon  (24 \epsilon -59)+24)}{12(2\e-1)}
\end{equation}
and also
\begin{equation}
\label{knavish}
C\du{\phi^4 \phi^4}{T}C\du{T\,\phi^l}{\phi^k} = 24 l (2-\e)^2 \delta_k^l\,.
\end{equation}
In which sense does this predict the leading behavior of the squared OPE coefficients? If we inverse Laplace transform the very leading coefficient, which is the term labelled $\#1$, then we find that
\be
\label{sumasympt}
\sum_{m: \D_m \leq \D_{\max} + d - \e + l \D_{\phi}}^{} C\du{\phi^4\, \phi^l}{m} C\du{\phi^4 m}{\phi^k} \sim 48 \delta_l^k \frac{\Delta_{\max}^{5 - 3 \epsilon}}{\Gamma(6 - 3\epsilon)} \xi(0)
\ee
For $k = l$ this estimate should be correct by the Tauberian theorem quoted above. This is confirmed for $\epsilon = 1$ and $l = k = 2$ by the top line in \cref{fig:sum_leading_beh}, where we plot the ratio $r$ between the left-hand side and just the leading term on the right-hand side -- the ratio converges to one as expected.

Let us now investigate the subleading terms. At $\epsilon = 1$, the first subleading terms arise from the terms labelled $\#2$ above. If we add the inverse Laplace transform of these terms on the right-hand side of \cref{sumasympt}, we find the improved convergence behaviour shown by the dashed line on the left in \cref{fig:sum_leading_beh}. Although this might look encouraging, the applicability of this result is limited. After all, we are actually not interested in the \emph{ratio} between the two terms but in rather in their \emph{difference} since we are trying to estimate the correct counterterm action. On the right in \cref{fig:sum_leading_beh} we see that the leading term $\#1$ subtracts a nice chunk of the value of the sum. We do even better by including $\#2$. However, at the next order we would run into trouble: there are visible subleading oscillations which do not decrease in size. Our approximations are based on keeping only a few terms in a power series in $\tau$, which (inverse) Laplace transform to a smooth function of $\Delta$. It is therefore simply not possible to reproduce such oscillating behaviour within our framework. We will address this issue below, but in the remainder of this subsection we will simply sidestep it by considering the summand itself.

To do so, let us first discuss how the summand
\be
C\du{\phi^4\, \phi^k}{m} C\du{\phi^4\, m}{\phi^l}|_{\Delta_m = \Delta}
\ee
admits a natural analytical continuation in $\Delta$. The OPE coefficients in \cref{opecoeffs1,opecoeffs2} are smooth functions of $n$, for fixed $k$ and $p$ . For each $p$, we thus obtain a smooth function of $n$, which we can trade for $\Delta_{m}$ (and hence $\Delta$) by simply setting $\Delta_m = (l + 4 - 2 p)\Delta_{\phi} + 2 n$. We then perform the finite sum over $p$ to obtain the desired continuation. We will evaluate this function (for now) at the shifted value $\Delta \to \Delta + d - \epsilon + k \Delta_\phi$ to take into account the extra $\tau$ dependence in the prefactor in \cref{schan}, just as we did in \cref{idcontr}.

The analytic continuation admits a simple asymptotic power series expansion. We find
\be
\label{eq:fullasymptoticsofsquaredopes}
\begin{split}
C\du{\phi^4\, \phi^l}{m} C\du{\phi^4\, m}{\phi^k}&|_{\Delta_m = \Delta + d - \epsilon + l \Delta_\phi} \\
= 48 \delta_l^k \Delta^{4 - 3\epsilon} \xi(0) & \left( \frac{1}{\Gamma (5-3 \epsilon )} +\frac{4 - 2 \epsilon}{\Gamma (4-3 \epsilon )} \Delta^{-1} + \frac{(2- \epsilon) (23 - 12 \epsilon)}{6 \Gamma (3-3 \epsilon )} \Delta^{-2} + \ldots \right)\\
+ 192 C\du{\phi^2 \phi^l}{\phi^k} \Delta^{2-2\epsilon} \xi(2-\epsilon) &\left( \frac{1}{\Gamma (3-2 \epsilon )} + \frac{4-2\e+\frac{(l-k)(2-\e) }{4}}{\Gamma (2-2 \epsilon )} \Delta^{-1} + \frac{\text{horrid}}{\Gamma(1-2\epsilon)} \Delta^{-2}  + \ldots \right)\\
+ 2 C\du{\phi^4 \phi^4}{T}C\du{T\,\phi^l}{\phi^k} \Delta^{-2 \epsilon} \xi(4-\epsilon) &\left(\frac{2 \e}{(3\e-4) \Gamma(1-2\epsilon)}+ \ldots  \right)\\
+ 144 C\du{\phi^4 \phi^l}{\phi^k} \xi(4 - 2 \epsilon) \Delta^{-\epsilon} &\left( \frac{1}{\Gamma(1 - \epsilon)} + \ldots\right)
\end{split}
\ee
As far as we checked, both the powers and coefficients in this expansion \emph{precisely} match those predicted by the inverse Laplace transform of the leading terms in $\FF\ud{\phi^k}{\phi^l}(\tau)$, with an extra factor $2$ arising only because the $\Delta$'s that contribute to the sum are spaced in units of two (since the Laplacian operator has dimension 2). In equations, we can say that to every order in the small $\tau$ and large $\Delta$ expansion:
\be
\mathcal L^{-1}\left[ S_d^{-2} \FF\ud{\phi^k}{\phi^l}(\tau); \tau \to \Delta \right] = \frac{1}{2} C\du{\phi^4\, \phi^l}{m} C\du{\phi^4\, m}{\phi^k}|_{\Delta_m = \Delta + d - \epsilon + l \Delta_\phi}
\ee
with the inverse Laplace transform obeying
\be
\mathcal L^{-1}\left[\tau^{-\rho};\tau \to \Delta\right] = \frac{\Delta^{\rho -1}}{\Gamma(\rho)}
\ee
and the right-hand side considered as an analytic function in $\Delta$. We have checked this claim for the identity operator (which reproduces the $\Delta^{4-3\epsilon}$ terms) to second subleading order, for the $\phi^2$ operator and the stress tensor (the $\Delta^{2 - 2\epsilon}$ and $\Delta^{-2\epsilon}$ terms) together also to second subleading order, and finally to leading order for the $\phi^4$ operator (the $\Delta^{-\epsilon}$ terms). Altogether, this shows that the subleading terms in the large $\Delta$ expansion would be captured perfectly by the subleading terms in $\FF_{ji}(\tau)$, were it not for the fact that we sum rather than integrate over the intermediate operators.

Ignoring (still) the issue of the oscillations, we can use the subleading terms in $\FF_{ji}(\tau)$ to compute the counterterms to high subleading orders. The counterterm Hamiltonian will then look like this:
\be
H_{\text{ct}} = g^2 S_d R^{2(d - \D_{\s})-1} \int_{S^{d-1}} d{\bf n} \left( \mathbf 1 X_{\mathbf 1} +  R^{\Delta_\phi^2} \phi^2 X_{\phi^2} + R^d T_{\tau \tau} X_{T} + R^{\Delta_\phi^4} \phi^4  X_{\phi^4}\right)
\ee
with dimensionless $X_{\op}$ coefficients which are roughly speaking determined by the inverse Laplace transform of $\FF_{ji}(\tau)$ given in \cref{eq:fullasymptoticsofsquaredopes}. If we set $\Lambda = \Delta_{\max}/R$, then it follows from a dimensional analysis argument that $X_\op \sim \Delta_{\max}^{2 \Delta_\s - d - \Delta_\op}$ to leading order. There are however several subtleties that we need to address before we can use \cref{eq:fullasymptoticsofsquaredopes} to obtain the explicit form of the $X_{\op}$. Let us discuss them one at a time.

\paragraph{The Offset in $\Delta_{\max}$} Our first subtlety is the extra prefactor $\exp\left(\tau (d - \epsilon + \Delta_i)\right)$ in \cref{schan} which leads to a small imperfection in the relation between the inverse Laplace transform of $\FF_{ji}(\tau)$ and the sum of squared OPE coefficients. This is reflected in the offset on the left-hand side of \cref{eq:fullasymptoticsofsquaredopes}. This will have to be taken into account by \emph{non-local} counterterms of the form\footnote{These operators can be made Hermitian by instead considering anti-commutators, for example $\{O_k,H_\text{CFT}\}$. The \emph{commutators} that this introduces are discussed further on.} already written down in \cite{PhysRevD.91.025005}:
\be
W \supset g^2 R^{2\e}\left(\frac{\D_{\max}}{R}\right)^{d-\D_k}\left( \int_{S^{d-1}_R} d{\mathbf n}\, \op_k \right) \left(\frac{ R H_{\text{CFT}}}{\D_{\max}} \right)^n
\ee
which leads to matrix elements of the form:
\be
R W\du{i}{j} \supset g^2 R^{2\e}S_d \D_{\max}^{d-\D_k-n}\D_i^n C\du{k i}{j}
\ee
Since $H_\text{CFT}$ is the integral of a local density but not a local operator itself, it follows that the counterterm is not local (unless $O_k=\textbf{1}$ and $n=1$). In practice, we get finitely many non-local counterterms because we keep only finitely many terms in the $\Delta_{\max}$ expansion.

\paragraph{The Denominator and the Integral} In our analysis we have seen that $\FF_{ji}(\tau)$ provides a good approximation of the sum of squared OPE coefficients $C\du{\sigma i}{k} C_{\sigma k j}$ but we need to still add the denominator $(\Delta_i - \Delta)$ in \cref{Hamevs} and integrate over $\Delta$. The $\Delta_i$ dependence in the denominator translates into yet another non-locality of the same type as dicussed in our previous point. Altogether, we can take into account both the offset and the denominator by the following replacement rule: any $\Delta^\alpha$ in the inverse Laplace transform of $\FF_{ji}(\tau)$ -- so in \cref{eq:fullasymptoticsofsquaredopes} -- needs to be replaced by
\be
\label{replDeltapowers}
\begin{split}
\D^\a \to \widehat{\D^{\a}_{\max}} &\colonequals \int^{\D_{\max}} d  \D  \frac{( \D - d + \e - R H_{\text{CFT}})^{\a}}{R H_{\text{CFT}} -  \D} \\
&= \D_{\max}^\a \left( - \frac{1}{\alpha} + \D_{\max}^{-1} \left( R H_{\text{CFT}} + \frac{\a (d - \e)}{\a - 1} \right) + O(\D_{\max}^{-2})\right)
\end{split}
\ee 
If we keep only finitely many terms in the $\D$-expansion, then the counterterm will be polynomial in $R H_\text{CFT}$.

\paragraph{$\Delta_{ij}$ Dependence in the Blocks} For non-identical operators, the conformal blocks depend on the difference $\Delta_{ij}$ in operator dimensions. This is a generic property, and in our case this shows up in the subleading terms in the expansion of $\FF_{ji}(\tau)$ given above in \cref{FjismalltauWF}. By their very nature, these terms only show up in the off-diagonal elements in sums like \cref{Hamevs}, so in the terms with $i \neq j$. These terms are unimportant for the computation of second-order energies, and therefore we can set to zero the $\Delta_{ij}$ terms in the inverse Laplace transform of the blocks. (See below for a more elaborate discussion.)

\paragraph{}

With this notation in place our counterterms take the form:
\be
\begin{split}\label{mainresult}
X_{\mathbf 1} &= -24  \, \xi(0)  \left( \frac{\widehat{\D_{\max}^{4 - 3\e}}}{\G (5-3 \e )} +\frac{(4 - 2 \e) \widehat{\D_{\max}^{3 - 3\e}}}{\G (4-3 \e )} + \frac{(2- \e) (23 - 12 \e) \widehat{\D_{\max}^{2 - 3\e}} }{6 \G (3-3 \e )} + \ldots \right)\\
X_{\phi^2} &= -96 \, \xi(2-\e) \left( \frac{\widehat{\D_{\max}^{2-2\e}}}{\G (3-2 \e )} + \frac{(4-2\e)\widehat{\D_{\max}^{1-2\e}}}{\G (2-2 \e )} + \frac{\text{horrid}\, \widehat{\D_{\max}^{-2\e}}}{\G(1-2\e)}  + \ldots \right)\\
X_{\mathbf T} &= - C\du{\phi^4 \phi^4}{T}\, \xi(4-\e) \left(\frac{2 \e \widehat{\D_{\max}^{-2 \e}} }{(3 \e -4 )\G(1-2\e)}+ \ldots  \right) \\
X_{\phi^4}&= -72 \, \xi(4 - 2 \e) \left( \frac{\widehat{ \D_{\max}^{-\epsilon}}}{\G(1 - \e)} + \ldots\right)
\end{split}
\ee
with all-important minus signs because they are supposed to \emph{cancel} divergences, and with $\Delta_{ij} \to 0$ also in `horrid' as given in \cref{horrid}. The preceding equation is the main result of this section, and few more comments are in order.

First of all, as noticed already in \cite{PhysRevD.91.025005}, the counterterm action has certain non-localities. Of course, a completely arbitrary non-local counterterm action would be worrying. However in this case, the non-locality enters in a mild and prescribed way, namely only through the substitution in \cref{replDeltapowers}. Unfortunately, at higher orders things appear less benign, for example in \cite{Elias-Miro:2017tup} a counterterm $:\int \phi^2 \int \phi^4 :$ was introduced. The restoration of locality in the continuuum limit therefore hinges on the irrelevance (in the technical sense) of these non-local counterterms. For the examples considered in the literature this seems to work well, but for less relevant (in the technical sense) perturbations this may become an issue.

Secondly, the conformal block decomposition of the four-point function organises the counterterms also in conformal multiplets. More precisely, we only need to add the conformal primary $\op_k(x)$ as an explicit operator in the counterterm Hamiltonian, and can then can take into account descendants (i.e. subleading terms in the block expansion) by improving its coefficient $X^k$. Notice that if we want finite energies at second order then we can extract the necessary counterterms from only those four-point functions $\FF_{ji}(\tau)$ with $j = i$, so the $\Delta_{ij}$ terms that appear in a general conformal block can be ignored.\footnote{We are again supposing here that all degeneracies have been resolved and therefore that $C\du{\s i}{j}$ has been diagonalised within small blocks as explained above.}

Thirdly, manifestly Lorentz-violating counterterms can only arise from other Lorentz-violating primary operators in the OPE expansion. Such counterterms are clearly \emph{allowed} as long as the vectorial indices point in the $\tau$ direction to preserve rotational invariance on the spacelike sphere, and then they are also \emph{expected} since they do not break more symmetries than our regulator.\footnote{The need for tensorial counterterms had been noticed in \cite{PhysRevD.91.025005}, but in the examples considered in that paper they were subleading and not worked out in detail.} Of course the integral of $T_{\tau \tau}$ is special since it is just the Hamiltonian again and despite appearances, it does not break Lorentz invariance. We believe that the corresponding counterterm can be interpreted as wave function renormalisation, which would be absent with a local Lorentz-invariant cutoff but does show up here. Notice also that it has the same matrix element as the subleading term proportional to $H_{\text{CFT}}$ in the expansion of the identity counterterm. However they appear with very different powers of $\Delta_{\max}$, so they are certainly different counterterms, and our analysis rigorously establishes the appearance of both.

Our fourth and last comment concerns the off-diagonal elements in \cref{Hamevs}. These terms are equally divergent and we would like to ask whether the counterterm action is expected to make the terms with $i \neq j$ finite as well. As we have discussed, this is not necessary to have renormalised energies at second order. In fact, because of the $\D_i$ in the denominator the sum in \cref{Hamevs} is not even Hermitian (real symmetric in this case) so there is no Hermitian counterterm that can make that expression finite to arbitrary subleading order. One may of course try to modify the expression to e.g.
\be
\label{symmetrizedsum}
\frac{V\du{i}{k} V\du{k}{j}}{\frac{1}{2}(\Delta_i + \Delta_j)- \Delta_k}
\ee
but the question then arises what the motivation would be for these ad hoc replacements.

An issue that is very much related to the issue of off-diagonal elements is the existence of counterterms that arise from commutator operators like
\be \label{offdiagct}
[H_{\text{CFT}}, \ldots[H_{\text{CFT}} , \op]] \sim \partial_t \ldots \partial_t \op
\ee
whose matrix elements between states $\langle j |$ and $|i\rangle$ are proportional to $\Delta_{ij}$, to an arbitrary power. These counterterms are compatible with the residual symmetries of the cylinder and so can in principle be added; one might in fact be tempted to do so at subleading orders when there is $\Delta_{ij}$ dependence in the blocks. However, to maintain a real symmetric Hamiltonian one can only add terms with even powers of $\Delta_{ij}$, and we have already seen that the blocks contain terms linear in $\Delta_{ij}$ above.

We believe the question of the off-diagonal term could be addressed by going one order higher and looking at the third-order correction. This is because, for a \emph{local} cutoff at least, we may expect the third-order counterterm to be completely local. If we look at \cref{3rdOrderEnergy}, this for example implies that divergences arising in the limit\footnote{More precisely, the analogue limit in the case of a local cutoff.} $j \to \infty$ for fixed but finite $k$ ought to be cancelled automatically by the counterterms in the latter two sums rather than by the third-order counterterm which is the last term in \cref{3rdOrderEnergy}. One could for example speculate that $W\du{i}{j}$ for a local cutoff instead makes finite an expression of the form
\begin{align}\label{3rdOrderMatices}
U\du{i}{j}&:=W\du{i}{i}+\frac{V\du{i}{k}V\du{k}{i}}{\D_{ik}}&\text{when }i=j\nonumber\\
U\du{i}{j}&:=W\du{i}{j}+\frac{V\du{i}{k}V\du{k}{j}}{2}\left(\frac{1}{\D_{ik}}+\frac{1}{\D_{jk}}-\frac{\D_i+\D_j}{\D_{ik}\D_{jk}}\right)&\text{when }i\neq j
\end{align}
Indeed, in terms of these new matrices, \cref{3rdOrderEnergy} becomes
\be
\frac{1}{R}\left(\frac{U\du{i}{j}V\du{j}{i}+V\du{i}{j}U\du{j}{i}}{\D_{ij}}+W\du{i}{i} \right)
\ee
and, provided the divergences take a local form, the diagonal elements of $W\du{i}{i}$ may now also be fixed to a local expression. 

Unfortunately, the TCSA cutoff is not local, so the reasoning of the previous paragraph does not obviously apply and it remains an interesting question to which extent we can use locality of third-order counterterms to gain insight in the off-diagonal terms at second order. For example, in \cite{Elias-Miro:2017xxf} non-local counterterms of the schematic form $\int \phi^2 \int \phi^4$ were introduced at third order, and it would be interesting to work out how second-order off-diagonal counterterms like those in \cref{offdiagct} would modify the coefficient of that counterterm.

\subsubsection{Oscillations and a Smoothly Varying Projector}\label{subsubsec:oscillations}
\cref{eq:fullasymptoticsofsquaredopes} relates the \emph{summand} in \cref{Xi} order-by-order to a simple asymptotic power series. The oscillations in \cref{fig:sum_leading_beh} exemplify how the full sum can only be replicated on average by an integral of the series. Up to now, we have flagrantly ignored these subleading oscillations, which arise inevitably from the discreteness of spectrum. One may wonder if there exists a more sophisticated counterterm action that takes into account these oscillations. For example, we can make the coefficients $X_\op$ in \cref{mainresult} more complicated functions of $\D_{\max}$ which also oscillate and then precisely cancel the oscillating sum. This is not a straightforward task and may drastically increase the non-locality of the counterterm action. In this section we therefore suggest how the issue can be mitigated by using a smoothed-out cutoff, centered around $\D_{\max}$. Concretely we will replace the TCSA projector $\mathcal{P}$ introduced above with a projector that smooths out the hard truncation at $\D_{\max}$. This leads to a new version of the replacement rule in  \cref{replDeltapowers}, but otherwise the counterterms are unchanged.

Concretely, we will consider a smoothly varying projector which flips from 1 to 0 around $\D_{\max}$ with a transition width growing slower than $\D_{\max}$. For example,
\be\label{smoothprojector}
\mathcal{P}\rightarrow \mathcal{P}_f:=\sum_{\D} \ket{\D}f(\D,\D_{\max})\bra{\D}
\ee
with
\be
f(\D,\D_{\max})\colonequals \frac{1}{2}\left(1-\tanh\left(\frac{\D-\D_{\max}}{\sqrt{\D_{\max}}}\right)\right)
\ee
Asymptotically, we have that (for $\a>0$)
\be
\int^\infty d\D \, f(\D,\D_{\max})\D^{\a-1}\sim \frac{f_\a(\D_{\max})\, \D_{\max}^\a
}{\a}
\ee
where
\be
f_\a(\D_{\max})\colonequals \sum_{k=0}^{\lfloor \a \rfloor}2(-1)^k (1-2^{2k-1})\left(\frac{\pi}{8}\right)^k \frac{B_{2k}}{(2k)!} \frac{\D_{\max}^{-k}}{(1+\a)_{-2k}}
\ee
and $B_n$ is the $n^\text{th}$ Bernoulli number. This implies that the reincarnation of \cref{replDeltapowers} with the smoothed cutoff should be
\begin{align}
	\D^\a \to \widehat{\widehat{\D^{\a}_{\max}}} &\colonequals \int^\infty d \D  \frac{(\D - d + \e - R H_{\text{CFT}})^{\a}}{R H_{\text{CFT}} - \D}f(\D,\D_{\max}) \\
	&= \D_{\max}^\a \left( - \frac{f_\a(\D_{\max})}{\a} + \frac{f_{\a-1}(\D_{\max})}{\D_{\max}} \left( R H_{\text{CFT}} + \frac{\a(d - \e)}{\a - 1} \right) + ...\right)\nonumber
\end{align}
Each singly hatted $\D_{\max}$ in \cref{mainresult} can be replaced with a doubly hatted one, and this gives the counterterms within the smoothly truncated theory.

\begin{figure}
\centering
\includegraphics[width=0.7\linewidth]{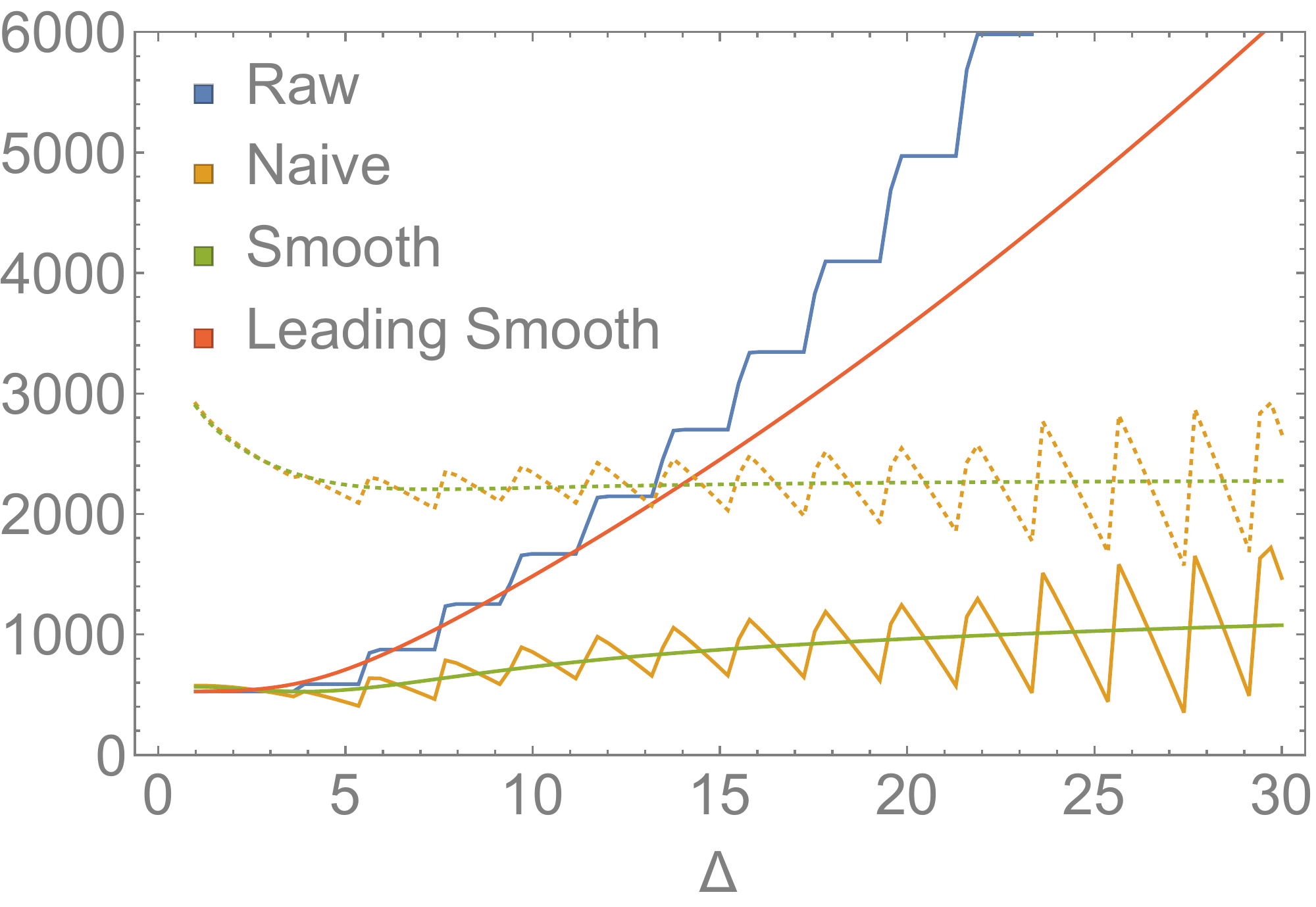}
\caption{The negative of \cref{divsum} in $\phi^4$ theory is plotted as a function of $\D_{\max}$ for $\op_i=\op_j=\phi^2$ and $\e=1/5$, so that $d=3.8$. The blue line is the raw data and the orange line incorporates the naive counterterms in \cref{mainresult}. The red line displays the same sum but with the smooth projector in \cref{smoothprojector}, and it includes the leading correction. Finally, the green line utilises all smooth counterterms. The dashed lines are discussed in \cref{decayingCTs}.}
\label{fig:smoothplot}
\end{figure}

\cref{fig:smoothplot} provides evidence that these smooth counterterms work in the $\phi^4$ theory. The oscillations that arose due to the discreteness of the spectrum get trampled and we end up with a convergent sum.

\subsubsection{Example: Three Dimensions}\label{sec:3d}
As an explicit example in an integer dimension, \cref{mainresult} with $\e=1$ becomes
\be
\begin{split}
	X_{\mathbf 1} &= -6\widehat{\D_{\max}}-12\widehat{\D_{\max}^0}\\
	X_{\phi^2} &= -48\widehat{\D_{\max}^0}
\end{split}
\ee
where each singly hatted $\D_{\max}$ is replaced with a doubly hatted one in the smooth theory.

Some care must be taken at this stage, since the replacement rules have to be extended to include logarithmic divergences:
\be
\begin{split}
	\widehat{\D_{\max}}&= -\D_{\max}+2\log \D_{\max}+\ldots\\
	\widehat{\D_{\max}^0}&= - \log \D_{\max}+\ldots
\end{split}
\ee
To this order, $\widehat{\widehat{\D_{\max}}}=\widehat{\D_{\max}}$ and $\widehat{\widehat{\D_{\max}^0}}=\widehat{\D_{\max}^0}$.

\begin{figure}
	\centering
	\includegraphics[width=0.7\linewidth]{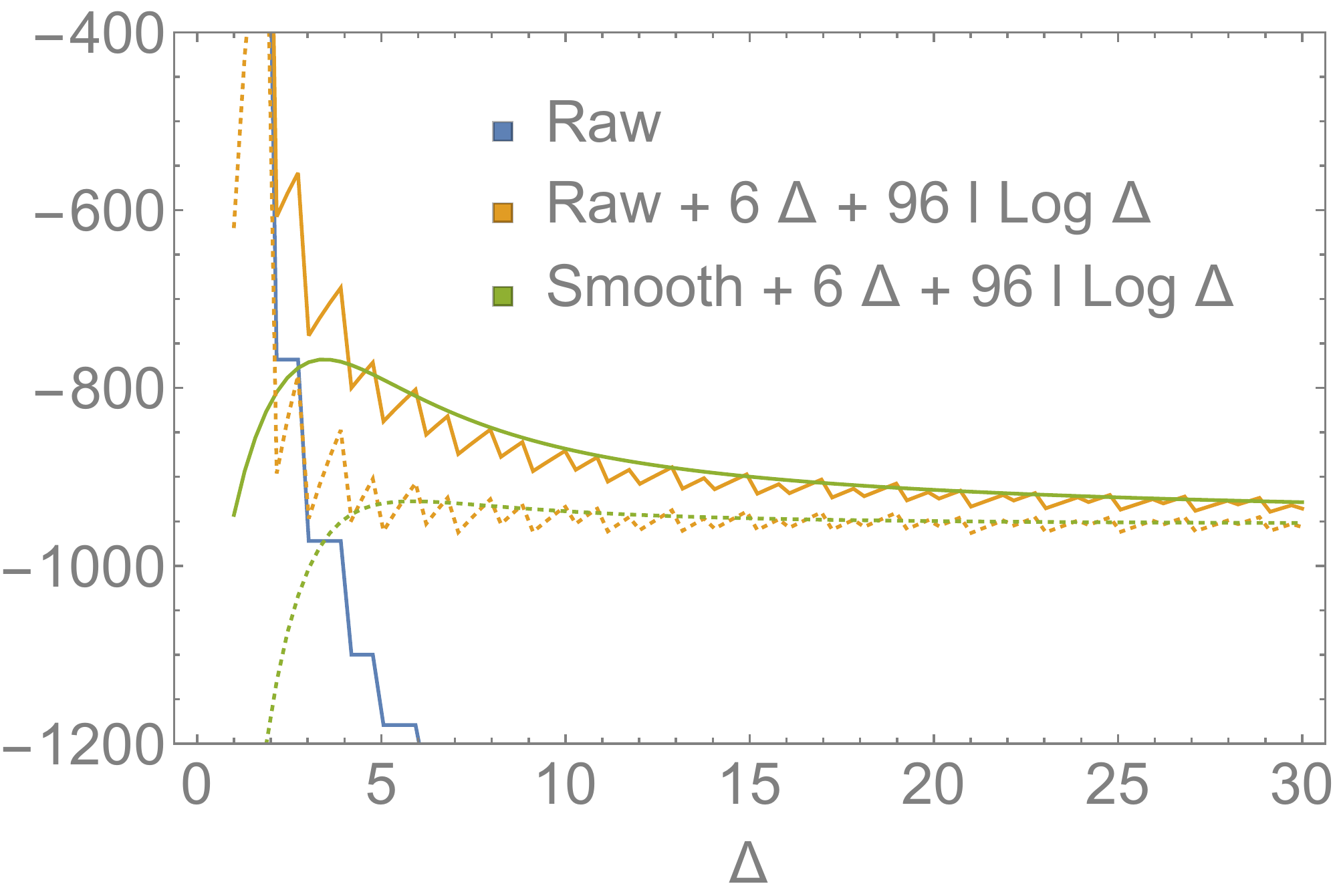}
	\caption{\cref{divsum} in $\phi^4$ theory is plotted as a function of $\D_{\max}$ for $\op_i=\op_j=\phi^2$ ($l=2$) and $\e=1$, so that $d=3$. The blue line is the raw data, whereas the orange line includes the addition of the naive counterterms and the green line is the regularised smoothed data. The oscillation amplitude appears to stay finite for this special case. The dashed lines are discussed in \cref{decayingCTs}.}
	\label{fig:epsilonequalsoneplot}
\end{figure}

\cref{fig:epsilonequalsoneplot} shows how these counterterms nicely regularise the $d=3$ variant of \cref{divsum}.

\subsubsection{Decaying Counterterms}\label{decayingCTs}
The counterterms in \cref{sec:3d} can be extended to include some decaying contributions, in an attempt to improve convergence. Taking the first decaying term,
\be
\begin{split}
	X_{\mathbf 1} &= -6\widehat{\D_{\max}}-12\widehat{\D_{\max}^0}\\
	X_{\phi^2} &= -48\widehat{\D_{\max}^0}\\
	X_{\phi^4} &= -36\widehat{\D_{\max}^{-1}}
\end{split}
\ee
where
\be
\begin{split}
	\widehat{\D_{\max}}&= -\D_{\max}+2\log \D_{\max}-2 R H_\text{CFT}\, \D_{\max}^{-1}+...\\
	\widehat{\D_{\max}^0}&= - \log \D_{\max}+R H_\text{CFT}\,\D_{\max}^{-1}+...\\
	\widehat{\D_{\max}^{-1}}&=\D_{\max}^{-1}+...
\end{split}
\ee
To this order, we have some discrepancy between singly and doubly hatted counterterms:
\be
\begin{split}
	\widehat{\D_{\max}}&=\widehat{\widehat{\D_{\max}}}+\frac{\pi^2}{12\,\D_{\max}}\\
	\widehat{\D_{\max}^0}&=\widehat{\widehat{\D_{\max}^0}}-\frac{\pi^2}{24\,\D_{\max}}
\end{split}
\ee
The dashed lines in \cref{fig:epsilonequalsoneplot} show how including these decaying counterterms can significantly improve convergence. Also note that the analogous dashed lines in \cref{fig:smoothplot}, which indicate a massive improvement in convergence that arises because the first correction is increasingly large as $\e\rightarrow 0$, that is, as $\phi^4$ becomes marginal.


\subsection{TCSA-Inspired Cutoffs for the Scalar Theory}\label{localcutoff}
The TCSA offers a natural and universally implementable cutoff. For the scalar theory we can however consider more customised cutoffs which would be suitable for perturbative computations. 

A particularly natural perturbative cutoff to consider would be a spatial momentum-space cutoff on the cylinder. In terms of canonical quantisation, this amounts to setting to zero all the creation operators with large momentum on the sphere. This would be a local cutoff in the sense that any momentum-space cutoff is local: it morally corresponds to discretising space to a lattice.\footnote{This cutoff would exactly correspond to a spatial lattice if the spatial manifold was a torus. For the sphere, we are not aware of an explicit lattice that would truncate the spherical harmonic expansion, but see \cite{Brower:2012vg} for an attempt at a lattice formulation of radial quantisation.} The state-operator correspondence maps these high-momentum modes to operators with many derivatives, and therefore it amounts to setting to zero operators with more than, say, $n_{\max}$ derivatives per $\phi$. This cutoff is however not easily implemented for our sum $\Xi\du{k\,l}{q}$ in \cref{Xi}, because we would then have to work out the Laplacians in the intermediate operator $\phi^{l-p} \square^n \phi^{k-p}$ and keep only those operators with less than $n_{\max}$ derivatives per $\phi$.

A somewhat related perturbative cutoff would be to limit the \emph{total} spatial momentum on the cylinder. Again through the state-operator correspondence, this cutoff amounts to a bound on the total number of derivatives in a give composite operator. For our sum $\Xi\du{k\,l}{q}$ given in \cref{Xi}, this truncation is simply:
\be \label{eq:spatialcutoff}
n \leq N_{\max}
\ee
and the UV cutoff scale is $\Lambda = 2 N_{\max} / R$. This cutoff does not appear to be local: for example, for a two-particle states the momentum of one particle is constrained in terms of the momentum of the other.

It is important to realise that the energy of a state (or the scaling dimension of an operator) can grow very large not only by taking large momentum (or many derivatives in the operator) but also by taking many particles (or many fundamental fields in the operator). Since we did not truncate the latter, any possible divergences arising from arbitrarily-many-particle states are not regulated by these cutoffs. Although this implies that these cutoffs are problematic non-perturbatively, such divergences are absent at any finite order in perturbation theory, since at every order a $\phi^k$ interaction adds only up to $k$ (i.e. finitely many) extra fields. This is why the momentum-cutoffs work only in perturbation theory.

Let us consider the $\phi^4$ theory again, now with the cutoff in total spatial momentum of \cref{eq:spatialcutoff}. Analysing the large $n$ behaviour of the summand is simple, both at the level of the integrand and using some simple modifications of the results in the previous section. The dimension of the intermediate operator in $\Xi\du{4\,l}{q}$ is given by
\be{}
\Delta = (l + k - 2 p) \Delta_\phi + 2n
\ee
and so to leading order the cutoff in $n$ essentially agrees with a cutoff in $\Delta$. We therefore propose the same leading-order counterterms. At subleading orders, there are small modifications. These are not interesting enough to write down explicitly, except that structurally we observe that with the cutoff of \cref{eq:spatialcutoff}:
\be \label{notsolocalcutoffasymptotics}
\begin{split}
{\Xi\du{k\,l}{q}} &= N_{\max}^{4 - 3\epsilon} (\# \delta_l^q  +  O(1/N_{\max}) )\\
&+ N_{\max}^{2 - 2 \epsilon} (\# C\du{\phi^2 \phi^l}{\phi^q}     +   \# (l - q + 6 - 12 \epsilon) N_{\max}^{-1} C\du{\phi^2 \phi^l}{\phi^q}  + O(1/N_{\max}^2))\\
&+ N_{\max}^{-\epsilon}  (\#C\du{\phi^4\phi^l}{\phi^q}   + \# (l - q - 4 - 4 \epsilon) N_{\max}^{-1} C\du{\phi^4\phi^l}{\phi^q} + O(1/N_{\max}^2))
\end{split}
\ee
where the coefficients $\#$ are unimportant functions of $\epsilon$ only.

We can offer two interesting observations about this cutoff. The first pertains to the oscillations discussed in the previous subsection. We observe that for the diagonal elements with $l = q$ the dependence on $l$ in \cref{notsolocalcutoffasymptotics} is fully captured by the OPE coefficients. This means that there exists non-trivial counterterm coefficients $X_\op$, functions of $\Delta_{\max}$ and $\epsilon$ only, which can get rid of the oscillations in $\Xi\du{k\,l}{q}$ completely. The question that arises now is whether this holds for arbitrary external operators -- in that case we could claim that this cutoff allows us to get rid of the oscillations without introducing drastic new non-localities.





Our second observation concerns the off-diagonal matrix elements. In \cref{notsolocalcutoffasymptotics}, we observe subleading off-diagonal terms, only one power down in the $N_{\max}$ expansion, with non-trivial dependence on $l$ and $q$, so the naive counterterms do not make the full matrix $\Xi\du{4\,l}{q}$ finite. In fact, as we pointed out before, the divergences are not Hermitian and so no reasonable counterterm action can cancel them.

Things marginally improve once we symmetrise the denominator as in \cref{symmetrizedsum}: in that case we find that \cref{notsolocalcutoffasymptotics} gets modified so that only even powers of $(l-q)$ appear, with the leading appearance one power further down. Remarkably, this is precisely the kind of subleading divergence that can in principle be addressed with Hermitian counterterms of the form $\del_t^2 \phi^2$ and $\del_t^2 \phi^4$. However, without a third-order analysis along the lines sketched in \cref{subsubsec:oscillations} we cannot be sure whether we need to add them.

Finally, passing to the more sophisticated \cref{3rdOrderMatices} we observe that, with the $N_{\max}$ cutoff, there is an additional non-trivial $l$-dependence in the analogue of \cref{notsolocalcutoffasymptotics}, and we need to either include non-local counterterms or new operators to cancel these terms. It follows that (with this cutoff) the naive counterterm action does not make \cref{3rdOrderMatices} finite. The subleading terms again have even powers of $(l-q)$ beginning at one power further down than in \cref{notsolocalcutoffasymptotics}.


\section{Conclusions}
The TCSA has proven to be a very useful numerical method for a wide variety of field theories. One of its main virtues is its simplicity, relying only on a simple Hamiltonian perspective that is familiar from quantum mechanics. In this work, we studied the TCSA cutoff in the framework of perturbative renormalisation and discussed some of its less attractive features like non-localities, non-covariant counterterms, and oscillations that are not easily cancelled. Fortunately, these effects are all suppressed by powers of the cutoff, and the suppression becomes stronger for more strongly relevant deformations. Morally speaking then, our results support the usual lore that the TCSA is at its most useful for strongly relevant deformations.

There are several possible future directions that naturally arise from this work. Firstly, we have seen that the conformal perturbation theory framework replaces the usual Feynman integrals with sums over computable free-field OPE coefficients. This offered us a remarkably simple way to compute second-order anomalous dimensions at the Wilson-Fisher fixed points. It would be interesting to see how this approach compares in difficulty to the Feynman diagram expansion in more general theories and/or at higher orders. How would the computational cost compare with the Feynman diagram expansion? Also, we have seen how the Hamiltonian viewpoint leads to an unconventional picture where anomalous dimensions arise from the diagonalisation of a single infinite matrix. We would be interested in learning the analogous picture for gauge theories and in particular integrable theories like planar $\mathcal N = 4$ SYM.

In numerical work, the focus has been on finding a counterterm action that approximates well the full matrix in \cref{Hamevs} (for reasons that are explained in \cite{PhysRevD.91.025005}). In perturbation theory we only care about diagonal elements in these kind of sums and there are subleading divergences in the off-diagonal terms that are not obviously cancelled by the counterterm action. In the future it would be interesting to understand these differences further. To do so one could for example work out the third-order correction for a local cutoff, which we expect to give us more information about the off-diagonal divergences at second order. Related to this, one could analyse the dependence of non-local divergences at third order (with the TCSA cutoff) on the details of the second-order counterterm action.

Finally, our work can provide a stepping stone for the numerical TCSA, in particular for the $\phi^4$ theories in three dimensions. We have provided the explicit second-order counterterm action in \cref{sec:3d} above. This counterterm action should suffice to get finite numerical results. However, to get \emph{accurate} results one may need to add further improvement terms, similar to those obtained in two dimensions in \cite{Elias-Miro:2017xxf,Elias-Miro:2017tup}, and perhaps a smoother cutoff like the one introduced above may be necessary to deal with any remaining oscillations. It would be very interesting to see if this will suffice to get accurate predictions from the TCSA in three spacetime dimensions.

\section*{Acknowledgments}
We would like to thank Matthijs Hogervorst and Slava Rychkov for discussions. DR is supported by an STFC studentship. BvR is supported in part by STFC under consolidated grants ST/P000371/1 and a grant from the Simons Foundation (\#488659).

\appendix

\section{Conformal Block Recursion Relations}\label{CBRR}
Conformal blocks obey a quadratic Casimir equation \cite{Dolan:2011dv}:
\begin{equation}
\mathcal{D} G_\D^{(l)}=\mathcal{C}_\D^{(l)}G_\D^{(l)}
\end{equation}
with the eigenvalue equal to
\begin{equation}
\mathcal{C}_\D^{(l)}=\D(\D-d)+l(l+d-2)
\end{equation}

The differential operator is given by
\begin{equation}
\frac{1}{2}\mathcal{D}=D_z+D_{\bar{z}}+(d-2)\frac{z\bar{z}}{z-\bar{z}}\Big((1-z)\partial_z-(1-\bar{z})\partial_{\bar{z}}\Big)
\end{equation}
where
\begin{equation}
D_z=z^2(1-z)\partial_z^2-(a+b+1)z^2\partial_z-abz
\end{equation}
and
\begin{equation}
a=-\frac{1}{2}\D_{12}\qquad b=\frac{1}{2}\D_{34}
\end{equation}

We can expand the conformal block as a sum of Gegenbauer polynomials \cite{PhysRevD.87.106004}:
\begin{equation}
G_\D^{(l)}=\sum_{n,m=0}^{\infty}c_{n,m}\mathcal{G}_\D^{(l)}(n,m)=|z|^\D\sum_{n,m=0}^{\infty}c_{n,m}|z|^n \frac{m!}{(2\n)_m}C_m^{\n}(\cos(\text{arg}(z)))
\end{equation}
where we have defined $\n=d/2-1$ and, in general, $c_{n,m}$ depends on $a,b,l,d$ and $\D$.

Defining
\begin{equation}
x=|z|\qquad y=\cos(\text{arg}(z))=\frac{z+\bar{z}}{2|z|}
\end{equation}
we have that
\begin{align}
\mathcal{D}&=\mathcal{D}_0+\mathcal{D}_1+\mathcal{D}_{\text{ext}}\\
\mathcal{D}_0&=x^2\partial^2_x-(1-y^2)\partial_y^2-(2\n+1)(x\partial_x-y\partial_y)\nonumber\\
\mathcal{D}_1&=x\left(-x^2 y \partial^2_x+y(1-y^2)\partial^2_y+2x(1-y^2)\partial_x\partial_y-xy\partial_x-(2\n+y^2)\partial_y\right)\nonumber\\
\mathcal{D}_{\text{ext}}&=-2x(a+b)(xy\partial_x-(1-y^2)\partial_y)-4xyab\nonumber
\end{align}

Using some Gegenbauer identities, we see that these operators act nicely on our summand: 
\begin{align}
\mathcal{D}_0\mathcal{G}_\D^{(l)}(n,m)&=\mathcal{C}_{\D+n}^{(m)}\mathcal{G}_\D^{(l)}(n,m)\\
\mathcal{D}_1\mathcal{G}_\D^{(l)}(n,m)&=-\g^{(+)}_{n,m}\mathcal{G}_\D^{(l)}(n+1,m+1)-\g^{(-)}_{n,m}\mathcal{G}_\D^{(l)}(n+1,m-1)\nonumber\\
\mathcal{D}_\text{ext}\mathcal{G}_\D^{(l)}(n,m)&=-\eta^{(+)}_{n,m}\mathcal{G}_\D^{(l)}(n+1,m+1)-\eta^{(-)}_{n,m}\mathcal{G}_\D^{(l)}(n+1,m-1)\nonumber
\end{align}
where
\begin{equation}
\g^{(+)}_{n,m}=\frac{(m+2\n)(\D+n+m)^2}{2(m+\n)}\qquad \g^{(-)}_{n,m}=\frac{m(\D+n-m-2\n)^2}{2(m+\n)}
\end{equation}
and
\begin{align}
\eta^{(+)}_{n,m}&=\frac{(m+2\n)\left((a+b)(\D+n+m)+2ab\right)}{m+\n}\\
\eta^{(-)}_{n,m}&=\frac{m\left((a+b)(\D+n-m-2\n)+2ab\right)}{m+\n}\nonumber
\end{align}
whence we find the following recursion relation:
\begin{equation}
c_{n,m}\left(\mathcal{C}_{\D+n}^{(m)}-\mathcal{C}_\D^{(l)}\right)=c_{n-1,m-1}\xi^{(+)}_{n-1,m-1}+c_{n-1,m+1}\xi^{(-)}_{n-1,m+1}
\end{equation}
where $\xi^{(+/-)}_{n,m}=\g^{(+/-)}_{n,m}+\eta^{(+/-)}_{n,m}$. With the initial condition $c_{0,m}=\d_{m,l}$, at the first level we have
\begin{align}
c_{1,l+1}&=\frac{l+2\n}{l+\n}\left(\frac{\D+l}{4}+\frac{a+b}{2}+\frac{ab}{\D+l}\right)\\
c_{1,l-1}&=\frac{l}{l+\n}\left(\frac{\D-l-2\n}{4}+\frac{a+b}{2}+\frac{ab}{\D-l-2\n}\right)\nonumber
\end{align}

Higher levels can then be found recursively. We note that the only non-zero coefficients at the second level are $c_{2,l},c_{2,l+2}$ and $c_{2,l-2}$.

\bibliography{../References}
\bibliographystyle{../utphys}
\end{document}